\documentclass[11pt,a4paper]{article}
\usepackage[margin=2.5cm]{geometry}
\usepackage{amsmath,amsfonts,amssymb}
\usepackage{graphicx}
\usepackage{authblk}
\usepackage{hyperref}
\usepackage{caption}
\usepackage{subcaption}
\usepackage{xspace}
\usepackage{float}
\usepackage{booktabs}
\usepackage{siunitx}
\usepackage{placeins}
\usepackage{makecell}
\usepackage[backend=biber, style=numeric, sorting=none]{biblatex}
\newcommand{\mgfive}{MG5aMC}

\title{\textbf{FPGA Acceleration of Matrix-Element Calculations for Monte Carlo Event Generation}} 

\author[*1]{H. Guti\'errez Arance}
\author[1]{F. Carri\'o} 
\author[1]{L. Fiorini}
\author[2]{S. Folgueras}
\author[1]{F. Herv\'as \'Alvarez}
\author[2]{P. Leguina L\'opez}
\author[1]{A. Oyanguren}
\author[1]{A. Valero}
\author[3]{C. Vico Villalba}

\affil[1]{Instituto de F\'isica Corpuscular (IFIC) - U. Valencia - CSIC}
\affil[2]{Universidad de Oviedo — ICTEA, Departamento de Física, C/Jesus Arias Velasco s/n, Oviedo}
\affil[3]{Rice University}
\affil[*]{Corresponding author: \texttt{hector.gutierrez@ific.uv.es}}

\date{}

\begin{document}
\maketitle

%\linenumbers

\begin{abstract}
We present an FPGA-based study of the acceleration of matrix-element
computations for Monte Carlo event generation, using
MadGraph5\_aMC@NLO as a benchmark framework. Two complementary benchmarks
are considered. First, we map the phase-space-generation and
matrix-element-evaluation stages of the benchmark process
\(e^+e^- \to \mu^+\mu^-\) onto an AMD Alveo U250 accelerator, enabling a
compute-only assessment of the mapped FPGA pipeline. Second, for the more
complex \(gg \to t\bar{t}+X\) processes with increasing jet multiplicity, we
investigate FPGA acceleration of the colour-algebra kernels as structured
targets for selective acceleration. In this second case, the reported
speedups correspond to the isolated colour-reduction kernel operating on
precomputed amplitudes, rather than to the full matrix-element evaluation or
the complete event-generation workflow.

The proposed implementations are developed using High-Level Synthesis and
are evaluated in terms of numerical accuracy, performance, device-level
energy efficiency, resource utilisation, and scalability. Under the stated
compute-only measurement methodology, the FPGA implementations achieve
substantial speedups and lower energy per event than the evaluated CPU and GPU
baselines. For the considered benchmarks, the numerical results show good
agreement on average with the corresponding CPU reference calculations, while
the resource analysis highlights the importance of numerical representation
in determining FPGA scalability. Taken together, these results demonstrate
the potential of FPGAs for accelerating selected matrix-element and
colour-algebra kernels, while further integrated evaluation is required to
determine their end-to-end benefit within heterogeneous Monte Carlo
event-generation workflows in high-energy physics.
\end{abstract}

\section{Introduction} %\input{introduction}
Due to the intrinsically probabilistic nature of fundamental particle
interactions and the composite structure of the proton, accurately modelling
proton collisions at the Large Hadron Collider (LHC) remains one of the major
challenges in high-energy physics (HEP). Reliable predictions require combining
computationally demanding perturbative quantum chromodynamics (QCD)
calculations of the hard-scattering process with non-perturbative inputs and
models, including proton parton distribution functions (PDFs) and
hadronisation models. Physics event generators provide the computational framework
required to combine these ingredients and obtain quantitative predictions for
a given theoretical model%
~\cite{Alwall:2014hca,HSFPhysicsEventGeneratorWG:2020gxw}.

For a partonic initial state \(i\), consisting of two incoming partons \(q_1\)
and \(q_2\), evolving into a final state \(X\), the hard-scattering contribution
is determined by the squared matrix-element. Schematically, the corresponding
partonic cross section can be written as
\begin{equation}
    \sigma(i \rightarrow X) \propto
    \int d\Phi\, |M_{i \rightarrow X}|^2,
    \label{eq:hard_scatter_bracket}
\end{equation}
where \(d\Phi\) denotes the corresponding phase-space measure.

To compute the cross section for a given process \(X\) at the LHC, the squared
matrix-element must be integrated over the accessible phase-space connecting
the initial and final states. In practice, the hard-scattering contribution is
calculated perturbatively up to a fixed order in an expansion in powers of the
relevant coupling constants. Physics event generators therefore rely on Monte
Carlo (MC) integration techniques, which approximate the integral by sampling
random points in the accessible phase-space. Modern MC generators provide
predictions at leading order (LO), while some also support higher-order
calculations such as next-to-leading order (NLO) and next-to-next-to-leading
order (NNLO)%
~\cite{Alwall:2014hca,Frederix:2018nkq}. These corrections are often
essential for precision phenomenology, especially given the ever-increasing
size of LHC datasets~\cite{HSFPhysicsEventGeneratorWG:2020gxw}. However,
increasing the perturbative order leads to a rapid and non-linear growth in
computational complexity, placing severe demands on computing
resources~\cite{HSFPhysicsEventGeneratorWG:2020gxw}.

The repeated evaluation of matrix-elements over large samples of independent
phase-space points makes Monte Carlo event generation particularly attractive
for hardware acceleration. This opportunity has already motivated significant
effort towards accelerating matrix-element integration and event generation on
vectorized CPU and GPU platforms%
~\cite{Carrazza:2021gpx,Valassi:2021ljk,Valassi:2023yud,
Hageboeck:2023blb,Valassi:2025xfn,Hageboeck:2025dpl}.

While CPUs and GPUs have already demonstrated the value of heterogeneous
computing for event generation%
~\cite{Carrazza:2021gpx,Valassi:2025xfn,Hageboeck:2025dpl},
the use of Field-Programmable Gate Arrays (FPGAs) in this domain remains
relatively unexplored%
~\cite{Barbone:2023mul,GutierrezArance:2025madgraphfpga}.
FPGAs enable custom hardware architectures tailored to specific computations,
offering fine-grained parallelism and deeply pipelined execution for
arithmetic-intensive workloads. They also provide an attractive
energy-efficiency profile, which is particularly relevant for large-scale HEP
computing, where power consumption is an increasingly important constraint.
Historically, FPGA programming has been challenging for complex algorithms
such as those arising in MC event generators, due to the large number of
arithmetic operations and the structured control flow associated with
Feynman-diagram calculations. Recent advances in high-level synthesis
(HLS)~\cite{AMD:VitisHLSUG1399} tools have significantly lowered these
barriers, making FPGA-based acceleration increasingly accessible%
~\cite{LeguinaLopez:2025hlshep,Duarte:2018ite}.

Previous work by the authors demonstrated the feasibility of porting
\mgfive-derived computations to FPGA platforms using high-level synthesis for
a simple benchmark process and a floating-point implementation%
~\cite{GutierrezArance:2025madgraphfpga}. In the present work, we extend that
proof of concept in two complementary directions: first, by introducing a
fixed-point FPGA implementation of the phase-space-generation and full
matrix-element-evaluation stages for \(e^+e^- \to \mu^+\mu^-\); and second,
by investigating kernel-level FPGA acceleration of colour-algebra calculations
for \(gg \to t\bar{t}+X\) processes with increasing jet multiplicity. Using
\textsc{MadGraph5\_aMC@NLO} as a benchmark framework and targeting an AMD
Alveo U250 accelerator~\cite{AMD:AlveoU250DS}, we present a comparative study
of numerical accuracy, throughput, device-level energy efficiency, resource
utilisation, and scaling behaviour across FPGA, CPU, and GPU platforms.

\section{MadGraph5\_aMC@NLO as a benchmark for hardware acceleration}

\textsc{MadGraph5\_aMC@NLO}~\cite{Alwall:2014hca} is one of the most widely used
frameworks for Monte Carlo event generation in high-energy physics. In recent
years, it has also incorporated significant acceleration-oriented developments
for GPUs and vector CPUs%
~\cite{Valassi:2021ljk,Valassi:2023yud,Hageboeck:2023blb,
Valassi:2025xfn,Hageboeck:2025dpl}. In particular, the CUDACPP plugin enables
the acceleration of LO processes on these platforms through extensive
refactoring for data-parallel execution, with a primary focus on
matrix-element calculations, which are traditionally one of the main
computational bottlenecks in event-generation workflows%
~\cite{Valassi:2025xfn}.

These developments have demonstrated substantial performance improvements,
although the speedup achieved depends on the process, numerical precision, and
hardware platform. GPU acceleration has been shown to yield matrix-element
speedups ranging from one to two orders of magnitude relative to
single-threaded Fortran implementations, with significant gains also observed
at the workflow level%
~\cite{Carrazza:2021gpx,Valassi:2025xfn,Hageboeck:2025dpl}.
At the same time, once the matrix-element calculation is accelerated, other
stages of the event-generation chain may become the limiting factor%
~\cite{Carrazza:2021gpx,Hageboeck:2025dpl}.

The \mgfive\ framework is particularly well suited as a benchmark for
hardware-acceleration studies because it combines physics flexibility with
automatic code generation~\cite{Alwall:2014hca}. Once the process definition,
model parameters, and integration settings are fixed, \mgfive\ produces
process-specific code for the corresponding matrix-element calculations. This
makes it possible to study metrics such as throughput, latency, numerical
accuracy, and resource usage starting from a realistic and widely used HEP
software stack, and therefore provides a natural benchmark framework for
assessing FPGA acceleration.

In this work, \mgfive\ is used as a benchmark framework in two complementary
ways. First, we consider the leptonic process
\(e^+e^- \to \mu^+\mu^-\), which provides a controlled baseline for mapping the phase-space-generation and matrix-element-evaluation stages
onto the FPGA. Second, we consider the more
complex hadronic processes \(gg \to t\bar{t}+X\), for which the size and
resource requirements of the generated matrix-element implementation increase
rapidly with final-state multiplicity. For the largest benchmark processes, the
complete matrix-element workflow could not be mapped onto the target FPGA
within the available resources. We therefore focus on the colour-algebra
kernels, whose regular arithmetic structure and increasing computational cost
make them suitable targets for studying selective FPGA acceleration.

Within \mgfive, the relevant event-generation workflow can be decomposed into
several stages, including random-number generation, phase-space sampling, and
matrix-element evaluation%
~\cite{Alwall:2014hca,Wettersten:2023ekm,Hageboeck:2025dpl}.
For complex multi-parton processes, matrix-element evaluation remains one of
the dominant computational stages, while the cost of the colour-algebra
calculation increases rapidly with the size of the colour basis. In the
representative \(gg \to t\bar{t}+2\)-jet workflow profiled in this study, the
isolated colour-algebra kernel accounts for slightly more than \(2\%\) of the
total execution time. For the higher-multiplicity
\(gg \to t\bar{t}+3\)-jet process, CPU profiling shows that colour summation
accounts for approximately \(7\%\) of the total execution time. This increase,
together with the rapid growth of the colour basis with multiplicity, provides
additional evidence of the increasing computational relevance of colour
algebra for more complex processes.

For the profiled two-jet workflow, accelerating the colour-algebra kernel
alone cannot provide a comparable end-to-end workflow speedup. Nevertheless,
its regular arithmetic structure and increasing computational cost with jet
multiplicity make it a useful and controlled starting point for evaluating
selective FPGA acceleration. This motivates the two benchmark strategies
adopted in this work: an accelerator-side implementation of the
phase-space-generation and full matrix-element-evaluation stages for a simple
benchmark process, and a kernel-level study of increasingly complex
\(gg \to t\bar{t}+X\) colour-algebra calculations.

\section{Methodology} 
This section presents the methodology used to assess FPGA-based acceleration
for selected stages of Monte Carlo event generation. The study combines two
complementary perspectives: an accelerator-side implementation of the
phase-space-generation and matrix-element-evaluation stages for the benchmark
process \(e^+e^- \to \mu^+\mu^-\), and a kernel-level acceleration study of
the colour-algebra kernels for \(gg \to t\bar{t}+X\) processes with increasing
jet multiplicity.

The target FPGA platform is an AMD Alveo U250 accelerator card designed for data-center and high-performance computing applications. Its numerical accuracy, performance, resource usage, and energy efficiency are evaluated by comparison with CPU and GPU implementations available within the \mgfive\ framework.

\subsection{Target Hardware Platform}
\label{sec:HW_plat}

The AMD Alveo U250 accelerator card is based on the Xilinx UltraScale+ XCU250 device~\cite{AMD:AlveoU250DS,AMD:AlveoU250UG}. Unlike system-on-chip FPGA platforms, the Alveo U250 operates as a PCIe-attached accelerator and does not include embedded hardware general-purpose processors. Instead, it is designed to accelerate compute kernels managed by a host CPU, making it well suited to heterogeneous high-performance computing workflows.

The XCU250 device provides 1,759,631 Look-Up Tables (LUTs), 3,660,140 Flip-Flops (FFs), 12,424 Digital Signal Processor (DSP) slices, and 5,442 Block RAM (BRAM) units. These resources enable the implementation of highly parallel and deeply pipelined hardware architectures, while the DSP blocks provide efficient support for numerically intensive arithmetic operations.

The platform also integrates multiple external DDR4 memory banks, with a total capacity of up to 64~GB, and supports PCIe Gen3 communication with the host system, enabling the handling of large data streams and intermediate buffers.

Overall, the combination of programmable logic, dedicated DSP resources, and external memory bandwidth makes the Alveo U250 a suitable platform for accelerating Monte Carlo event-generation kernels, particularly in dataflow-oriented designs aimed at high-throughput execution.

\subsection{Event Implementation}
\label{sec:event_implementation}

In matrix-element generators such as \mgfive, the evaluation of each event follows a well-defined computational workflow. First, a phase-space point is sampled and a valid external momentum configuration is generated for the particles in the process. Then, for each phase-space point, the corresponding matrix-element is evaluated and used to determine the event weight. Although these steps are traditionally executed sequentially for a single event, they are naturally parallel across large samples of independent phase-space points~\cite{Wettersten:2023ekm}.

\begin{figure*}[t]
    \centering
    \includegraphics[width=\textwidth]{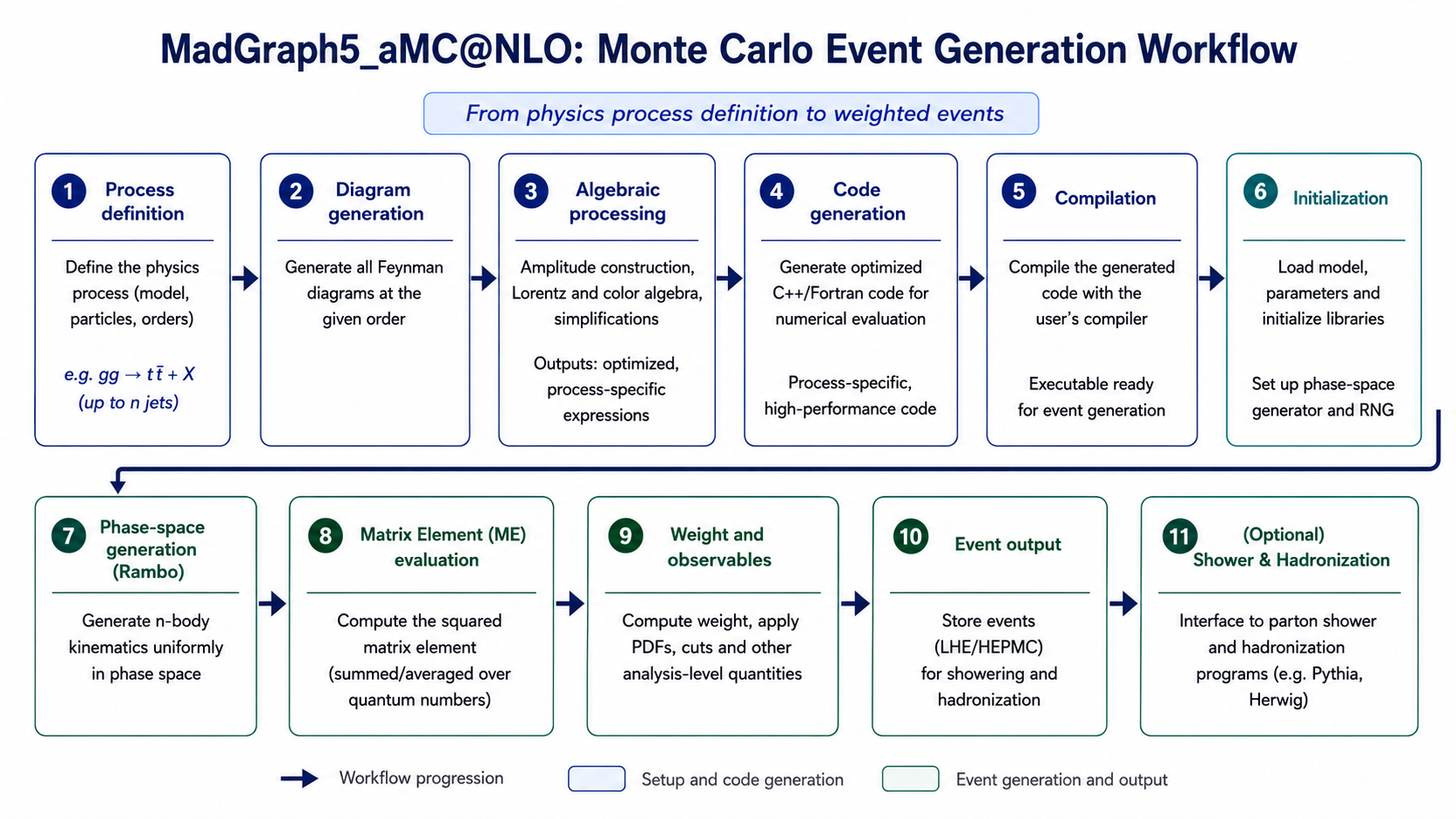}
    \caption{General \mgfive\ workflow for Monte Carlo event generation, from process definition and diagram generation to phase-space generation, matrix-element evaluation, and event output.}
    \label{fig:madgraph_workflow}
\end{figure*}

Figure~\ref{fig:madgraph_workflow} illustrates the general event-generation workflow in \mgfive, from process definition and diagram generation to phase-space sampling, matrix-element evaluation, and event output. This workflow provides the conceptual framework for the implementation strategies considered in this work.

We focus on the stages of this workflow that are most relevant for FPGA acceleration. In particular, matrix-element evaluation constitutes the dominant computational cost, while colour-algebra becomes an increasingly important component for more complex multi-parton processes. The following subsections describe how these stages are implemented and mapped onto the target FPGA platform.

\subsubsection{Phase-space sampling and momentum generation}

The generation of phase-space points is the first step in the event-generation workflow. It consists of constructing valid kinematic configurations for the external particles from pseudo-random inputs, subject to energy--momentum conservation. In this work, we employ a RAMBO-based~\cite{Kleiss:1985gy} phase-space generator to produce the corresponding sets of four-momenta.

Each phase-space point defines a complete external momentum configuration and serves as input to matrix-element evaluation. In the present implementation, pseudo-random number generation is performed on the host system, while the phase-space points are constructed directly on the FPGA accelerator. This enables a streaming execution model in which random inputs are transferred from the host and the full kinematic configuration is generated on-device.

Although the computational cost of phase-space generation is much smaller than that of matrix-element evaluation, it remains an essential component of the overall workflow. Integrating this stage within the FPGA pipeline reduces data-movement overheads and allows the event-generation chain to be organized as a continuous streaming process.

\subsubsection{Matrix-Element computation}

The evaluation of the matrix-element consists of computing the squared
scattering amplitude for a given phase-space point, defined by the
four-momenta of the external particles. Within the \mgfive\ framework, this
calculation is performed by evaluating the relevant Feynman-diagram
contributions and summing them over the allowed helicity and colour
configurations~\cite{Alwall:2014hca}.

At the computational level, the HELAS formalism represents external particles
through numerical wavefunctions and combines them through interaction vertices
to construct intermediate wavefunctions and diagram amplitudes. Particle
couplings and propagator factors are applied during this construction, after
which the resulting complex diagram amplitudes are coherently combined for
each helicity configuration. The colour-decomposed amplitudes are subsequently
contracted with the colour matrix, as described in the following subsection.
The generated computation therefore consists mainly of real and complex
additions and multiplications arranged according to the dependency graph of
the corresponding Feynman diagrams.

The construction of intermediate wavefunctions is recursive only in the sense
that each wavefunction depends on quantities evaluated at earlier stages. Once
the process-specific code has been generated, however, the complete dependency
graph and execution order are fixed. The hardware therefore evaluates a static
and deterministic sequence of arithmetic operations, without dynamic recursion
or data-dependent control flow.

Parallelism is applied differently at the helicity, operation, and event
levels. The loop over the allowed helicity configurations is fully unrolled,
exposing the calculations associated with the different helicity
configurations as concurrent hardware paths. The corresponding contributions
for a given event can therefore be evaluated in parallel and subsequently
combined through a fixed reduction structure.

Within each helicity path, the complete sequence of HELAS operations is not
executed simultaneously. External wavefunctions, intermediate wavefunctions,
and vertex amplitudes are connected by data dependencies that must be
preserved. Operations whose inputs are available independently can be
scheduled concurrently, whereas dependent operations are mapped to successive
stages of the arithmetic datapath. The resulting architecture therefore
exploits the operation-level parallelism available in the generated dependency
graph while using pipelining to overlap the execution of its successive
stages.

A further level of concurrency is obtained across independent phase-space
points. The event-processing loop is pipelined with an initiation interval of
one clock cycle for the \(e^+e^- \to \mu^+\mu^-\) implementation. Once the
pipeline has been filled, a new phase-space point can be accepted every clock
cycle while previously accepted events continue through later stages. This is
possible because the matrix-element evaluations of different phase-space
points do not exchange information. The architecture thus combines
helicity-level spatial parallelism, dependency-aware operation-level
pipelining, and event-level streaming concurrency.

The size of the generated arithmetic graph depends on the physics process,
including the number of diagrams, helicity configurations, intermediate
wavefunctions, and colour structures. The overall computational cost of the
complete matrix-element evaluation is therefore process dependent. A
quantitative complexity analysis is provided for the colour-algebra
contraction considered in the kernel-level benchmarks, for which the operation
count can be derived directly from the size of the colour basis.

In this work, the matrix-element computation is mapped onto the AMD Alveo U250
using a custom streaming architecture. Phase-space points generated on the
FPGA are fed directly into the computation pipeline, where the different
stages operate concurrently. This organization provides a predictable latency
and sustained streaming throughput. For the simple benchmark process, the phase-space-generation and matrix-element-evaluation stages can be
mapped onto a single FPGA pipeline. For the more
complex multi-parton processes, however, the size and resource requirements of
the generated arithmetic graph grow substantially, motivating the selective
acceleration strategy described in the following subsection.

\subsubsection{Colour-algebra calculation}

For the gluon-initiated \(gg \to t\bar{t}+X\) processes considered in this
work, one of the final reduction steps in matrix-element evaluation consists
of combining the partial amplitudes associated with different colour
configurations. In the \mgfive\ framework, this is formulated through a
colour-flow decomposition, in which the total amplitude is expressed in terms
of colour-flow components that are subsequently contracted with a colour
matrix~\cite{Alwall:2014hca}.

More specifically, the squared matrix-element can be written as
\begin{equation}
\label{eq:color}
|\mathcal{M}|^2 =
\sum_{i,j}
\mathcal{A}_i^* \, C_{ij} \, \mathcal{A}_j ,
\end{equation}
where \(\mathcal{A}_i\) denote the colour-decomposed amplitudes and
\(C_{ij}\) is the corresponding colour matrix. This operation performs the
contraction over the colour basis and constitutes a structured reduction step
within the matrix-element calculation.

Let \(B\) denote the number of colour-decomposed amplitudes. A direct dense
evaluation of Eq.~\ref{eq:color} considers \(B^2\) amplitude pairs
\((i,j)\), and its computational complexity therefore scales as
\(\mathcal{O}(B^2)\) with the size of the colour basis. In the FPGA
implementation used in this work, the real and symmetric structure of the
colour matrix is exploited through a folded triangular formulation. The
diagonal and upper-triangular matrix positions are traversed once, while the
precomputed coefficients account for the corresponding symmetric
contributions. The resulting traversal contains
\(B(B+1)/2\) matrix positions, reducing the number of explicitly evaluated
pairs while preserving the quadratic asymptotic complexity.

For the \(gg \to t\bar{t}+1\) jet,
\(gg \to t\bar{t}+2\) jets, and
\(gg \to t\bar{t}+3\) jets benchmark kernels, the colour bases contain
6, 24, and 120 amplitudes, respectively. In the fundamental and colour-flow
bases, the size of the colour matrix is known to grow factorially with
particle multiplicity, leading to a correspondingly rapid, typically
factorial-squared, growth of the colour-algebra cost~\cite{Lifson:2022eti}.
The increase observed in the benchmark processes considered here is
consistent with this behaviour. In the folded triangular formulation used in
our FPGA implementation, the 6, 24, and 120 amplitudes correspond to 21, 300,
and 7260 explicitly traversed matrix positions, respectively.

Thus, increasing jet multiplicity is accompanied by a substantial increase in
the computational and storage requirements of the colour contraction. This
complexity analysis applies specifically to the colour-reduction kernel; the
computational cost of the complete matrix-element calculation also depends on
process-specific quantities such as the number of Feynman diagrams, helicity
configurations, and intermediate wavefunctions.

From a hardware perspective, the colour contraction consists of repeated
complex products and accumulations with a regular data-access pattern. This
structure can be mapped onto parallel and pipelined arithmetic units more
directly than the broader matrix-element pipeline, whose wavefunction and
amplitude evaluations follow a larger process-specific dependency graph.

\begin{figure*}[t]
    \centering
    \includegraphics[width=\textwidth]{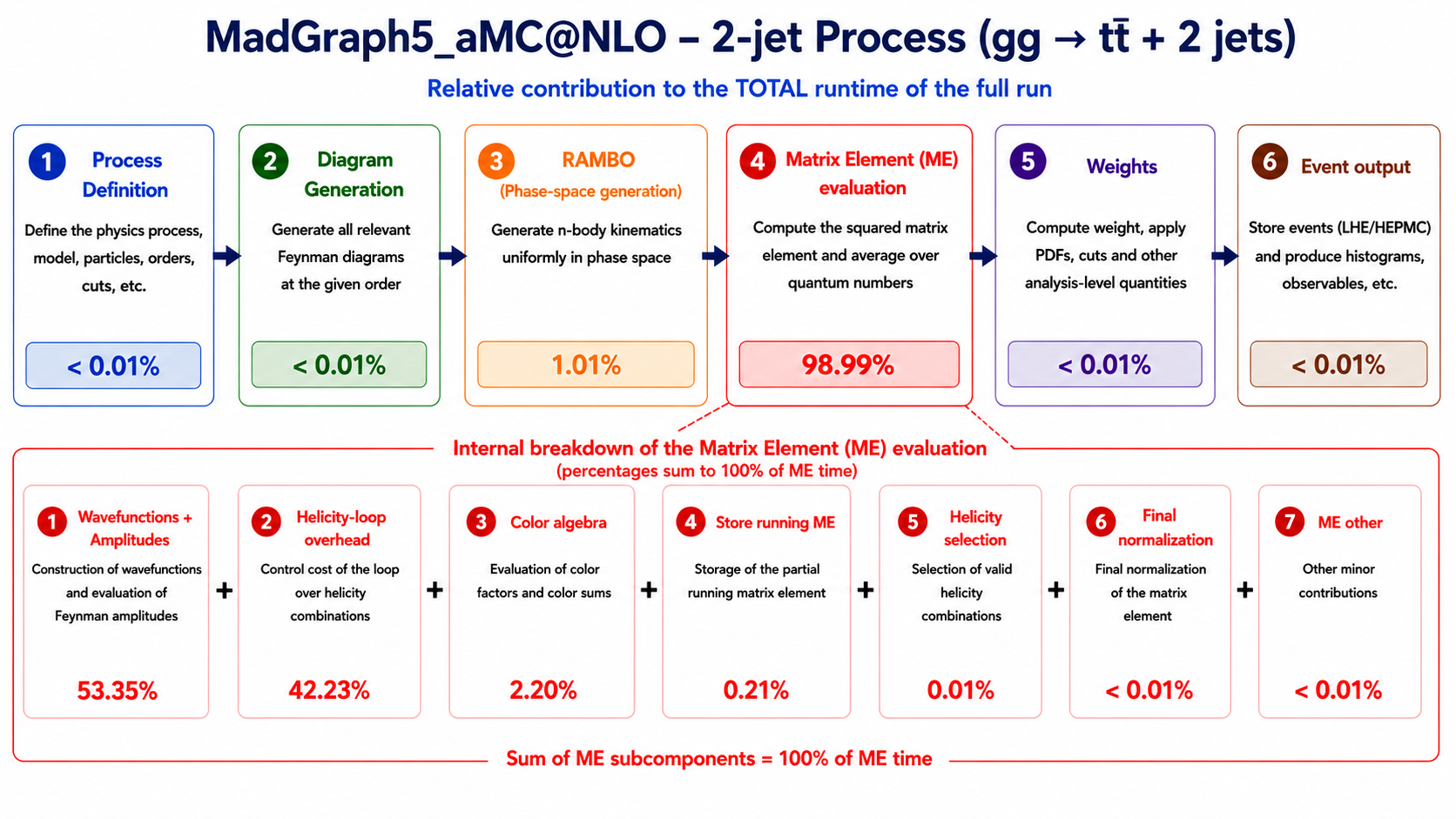}
    \caption{Execution-time breakdown for the
    \(gg \to t\bar{t}+2\) jets process. The top row shows the relative
    contribution of the main workflow stages to the total runtime, while the
    bottom row details the internal breakdown of the matrix-element
    evaluation.}
    \label{fig:ggttgg_breakdown}
\end{figure*}

Figure~\ref{fig:ggttgg_breakdown} shows the execution-time breakdown for the
representative \(gg \to t\bar{t}+2\) jets process. Matrix-element evaluation
dominates the total runtime, while the colour-related stages represent only a
relatively small fraction of the full computation for this benchmark. In this
case, the colour-reduction stage accounts for slightly more than \(2\%\) of
the total execution time.

For the higher-multiplicity \(gg \to t\bar{t}+3\) jets process, CPU profiling
shows that the colour-summation stage accounts for approximately \(7\%\) of
the total execution time. The increase from slightly more than \(2\%\) to
approximately \(7\%\), together with the rapid growth of the colour-basis
size, provides additional quantitative evidence of the increasing
computational relevance of colour algebra at higher multiplicity.

Nevertheless, even a substantial acceleration of the isolated colour
contraction should not be interpreted as an equivalent end-to-end workflow
speedup. Its regular structure and its increasing cost with the colour-basis
size make it a useful controlled target for selective FPGA acceleration.
Previous work has also identified colour computation in \mgfive\ as a relevant
target for dedicated optimisation in high-multiplicity QCD
processes~\cite{Lifson:2022eti}.

Accordingly, the colour-reduction stage is selected as the kernel-level FPGA
target for the \(gg \to t\bar{t}+X\) benchmarks. The corresponding
host--FPGA partitioning, hardware implementation, and evaluation scope are
described in the following FPGA-implementation subsection.

\subsection{FPGA implementation}

The computational stages described in
Section~\ref{sec:event_implementation} are mapped onto the FPGA to exploit
their inherent parallelism. For the benchmark process
\(e^+e^- \to \mu^+\mu^-\), the phase-space-generation and
matrix-element-evaluation stages are implemented as a single accelerator-side
pipeline. The pipeline consumes pseudo-random inputs prepared by the host and
produces the corresponding matrix-element results. For the more complex
\(gg \to t\bar{t}+X\) processes, a kernel-level acceleration strategy is
adopted, focusing on the colour-algebra component as a structured computational
stage whose cost increases with jet multiplicity.

Together, these two approaches make it possible to study how resource utilization, numerical representation, and achievable performance evolve as process complexity increases. All FPGA kernels are developed using an HLS flow and integrated within the Xilinx Vitis toolchain.

\subsubsection{Overall architecture}

The computational workflow described in Section~\ref{sec:event_implementation} is implemented on a heterogeneous architecture composed of a host CPU and an FPGA accelerator, namely the AMD Alveo U250~\cite{AMD:AlveoU250DS,AMD:AlveoU250UG}. The host system is responsible for runtime control, pseudo-random input preparation, and result retrieval, while the FPGA executes the core computational stages of the event-evaluation pipeline.

Data transfer between the host and the accelerator is performed through a PCIe Gen3 interface. Input data are first written to global memory (DDR4) accessible by the FPGA, from which they are consumed by the hardware kernels. The FPGA executes the subsequent stages of the computation on-device and writes the resulting observables back to global memory for retrieval by the host.

\begin{figure*}[t]
    \centering
    \includegraphics[width=\textwidth]{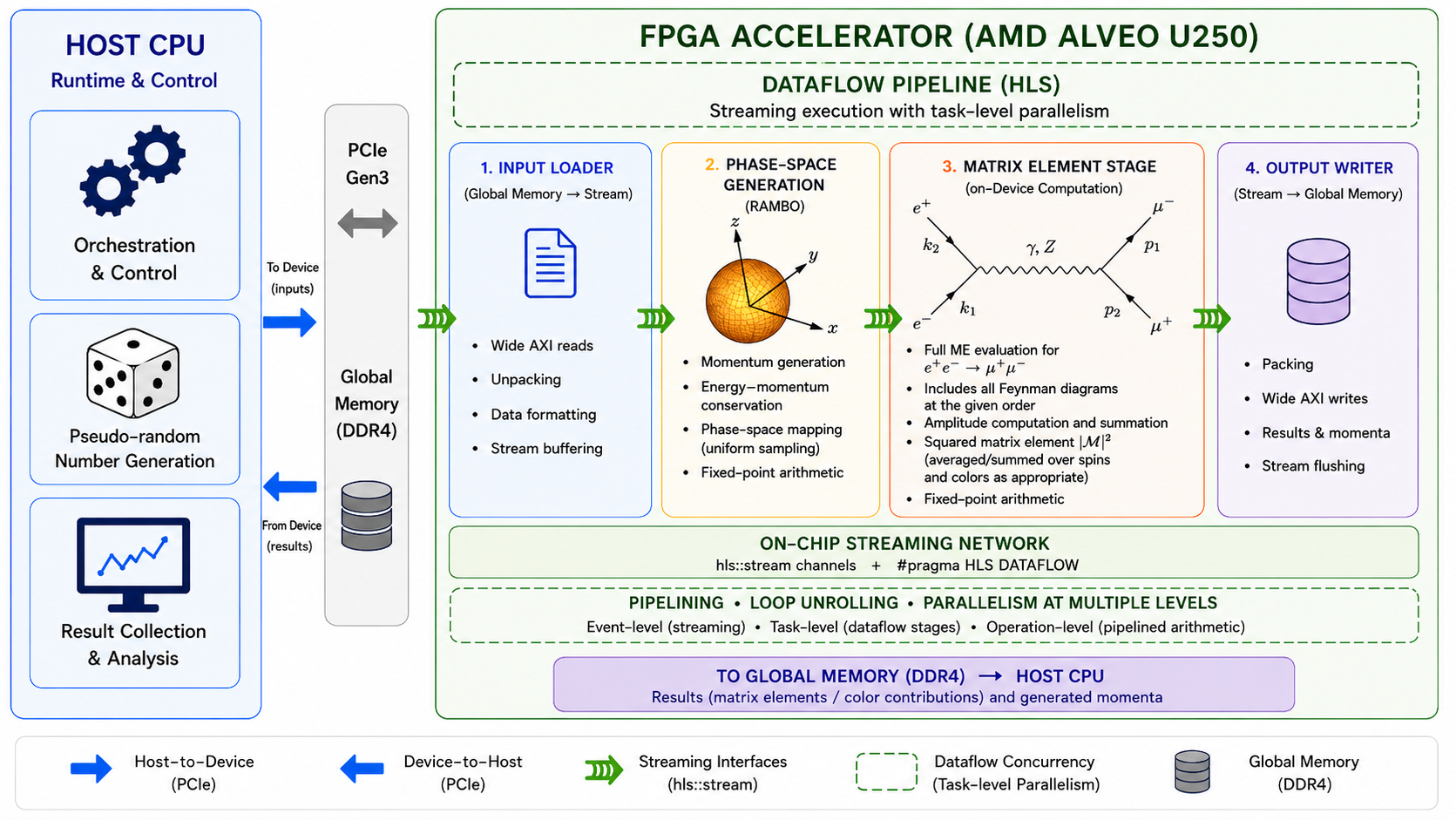}
    \caption{Overview of the heterogeneous host--FPGA architecture adopted in this work. The host CPU manages runtime control, input preparation, and result retrieval, while the AMD Alveo U250 executes the streaming computational pipeline. Depending on the target workflow, the compute stage implements either the full matrix-element evaluation or the colour-algebra kernel.}
    \label{fig:fpga_architecture}
\end{figure*}

The FPGA kernel follows a streaming dataflow architecture, as illustrated in Figure~\ref{fig:fpga_architecture}. The computation is decomposed into a sequence of processing stages connected through on-chip streaming channels (\texttt{hls::stream}), which are executed concurrently through the use of the \texttt{\#pragma HLS DATAFLOW} directive.

The overall pipeline consists of four main stages:
\begin{itemize}
    \item \textbf{Input loader}, which reads data from global memory and formats them into streaming packets.
    \item \textbf{Phase-space generation}, where particle momenta are constructed using a RAMBO-based algorithm~\cite{Kleiss:1985gy}.
    \item \textbf{Compute stage}, which performs either the full matrix-element evaluation or the colour-algebra kernel depending on the target workflow.
    \item \textbf{Output writer}, which packs the results and writes them back to global memory.
\end{itemize}

Each stage is internally pipelined to sustain high-throughput execution, while the streaming interfaces allow continuous data propagation across the pipeline. This architecture provides a unified execution model for both implementation strategies considered in this work: full event evaluation for simple processes, and selective acceleration of the colour-algebra stage for more complex ones.

\subsubsection{Full matrix-element implementation for
\(e^+e^- \to \mu^+\mu^-\)}

For the benchmark process \(e^+e^- \to \mu^+\mu^-\), the
phase-space-generation and full matrix-element-evaluation stages are
implemented on the FPGA as a single accelerator-side pipeline. The pipeline
consumes pseudo-random inputs prepared by the host, constructs the external
four-momenta, and evaluates the corresponding matrix element on-device. The
results are subsequently written to global memory for retrieval by the host.

The implementation is organised into four main processing stages. First,
host-generated pseudo-random values are read from global memory through an AXI
interface and unpacked into fixed-size data packets. These packets are then
forwarded through on-chip streaming channels to the phase-space-generation
stage.

In the second stage, phase-space momenta are generated directly on the FPGA
using a RAMBO-based algorithm adapted for hardware execution. The
implementation operates on fixed-point data types and computes the
four-momenta of the external particles while enforcing energy--momentum
conservation in the centre-of-mass frame. The resulting momenta are streamed
directly to the matrix-element stage without intermediate storage in external
memory.

The third stage performs the matrix-element evaluation through a hardware
implementation of the HELAS formalism~\cite{Murayama:1992gi}. External and
intermediate wavefunctions, propagators, and interaction vertices are
evaluated using fixed-point arithmetic and complex data types. The amplitudes
associated with the individual Feynman diagrams are then computed and
coherently combined to obtain the total scattering amplitude for each
helicity configuration.

The summation over helicity configurations is fully unrolled, exposing the
corresponding calculations as concurrent hardware paths. For each
configuration, the squared amplitude is computed from the real and imaginary
components of the complex amplitude. The resulting contributions are
accumulated using a wider fixed-point type to provide additional
dynamic-range headroom and reduce the risk of overflow. The final
matrix-element value is obtained after averaging over the initial-state
helicities.

In the final stage, the computed results are converted to the external output
representation, packed into memory words, and written to global memory for
subsequent retrieval by the host.

Overall, the design exploits event-level parallelism through the continuous
processing of independent phase-space points, together with operation-level
parallelism exposed through pipelining and loop unrolling within the
individual stages. The implementation uses fixed-point arithmetic with
explicit scaling of dimensional quantities, including particle momenta,
masses, and widths. This representation was adopted to limit resource
pressure while retaining the numerical behaviour evaluated in
Section~\ref{sec:numerical_validation_full}. The scalability of this approach
is nevertheless constrained by the increasing size of the generated
arithmetic graph for processes involving larger numbers of Feynman diagrams,
helicity configurations, and non-trivial colour structures.

\subsubsection{Colour-algebra acceleration for \(gg \to t\bar{t}+X\) processes}

For more complex processes, such as \(gg \to t\bar{t}+X\), mapping the full
matrix-element computation onto the FPGA becomes increasingly constrained by
resource usage and arithmetic complexity. For this reason, a selective
acceleration strategy is adopted, focusing on the colour-algebra component as
a structured part of the overall workflow.

\begin{figure*}[t]
    \centering
    \includegraphics[width=\textwidth]{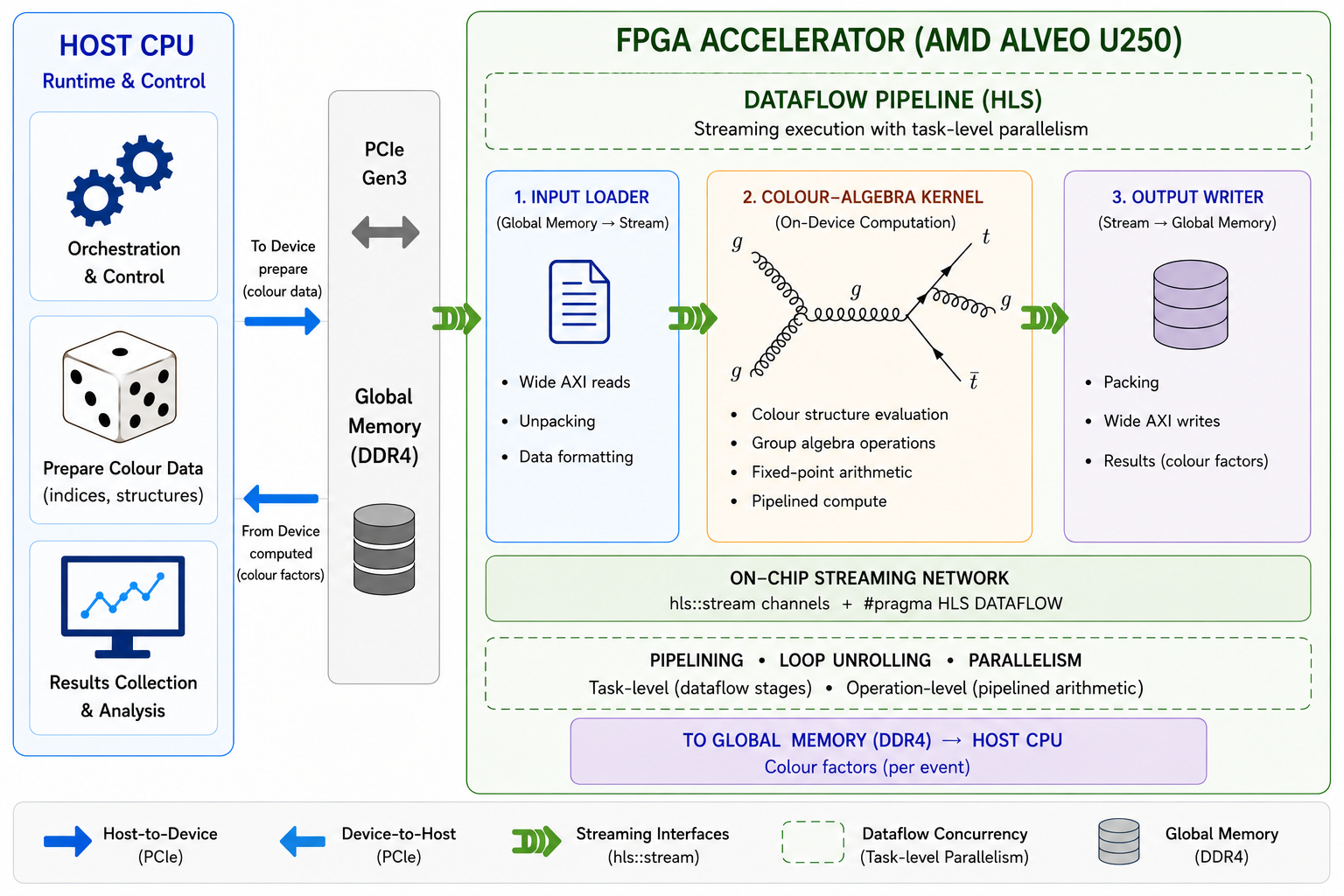}
    \caption{Overview of the selective acceleration strategy used for the
    colour-algebra kernels. The host CPU prepares the precomputed
    \texttt{jamp} amplitudes and transfers them to the AMD Alveo U250 through
    PCIe and global memory. The FPGA executes a streaming dataflow pipeline
    composed of an input loader, the colour-reduction kernel, and an output
    writer. The resulting colour factors are written back to global memory
    and retrieved by the host.}
    \label{fig:color_architecture}
\end{figure*}

In this approach, the FPGA does not evaluate the full set of wavefunctions,
propagators, and interaction vertices. Instead, the precomputed
\texttt{jamp} amplitudes associated with the different colour configurations
are transferred to the FPGA, where the colour-reduction kernel is evaluated.
The accelerator therefore computes the contraction of these amplitudes with
the corresponding colour matrix, yielding the colour contribution to the
squared matrix-element. The corresponding execution flow is illustrated in
Figure~\ref{fig:color_architecture}.

From a computational perspective, the colour-algebra corresponds to a
matrix--vector contraction followed by a reduction over colour indices,
\begin{equation}
\label{eq:delta_me}
\Delta \mathrm{ME} = \sum_i A_i^* \sum_j C_{ij} A_j ,
\end{equation}
where \(A_i\) are the colour-flow amplitudes, \(C_{ij}\) is the colour
matrix, and \(\Delta \mathrm{ME}\) denotes the colour-reduction contribution
evaluated by the FPGA kernel. For the FPGA implementation, this contraction is
evaluated using a folded triangular form that is analytically equivalent to
the full double sum used in \mgfive. This reformulation exploits the fact that
the colour matrix is real and symmetric: diagonal terms are treated
explicitly, while off-diagonal contributions are traversed only once through
the upper-triangular part of a precomputed normalized matrix, whose
coefficients already include the corresponding symmetric factor.

Let \(B\) denote the number of colour-flow amplitudes. A direct dense
evaluation of Eq.~\ref{eq:delta_me} considers \(B^2\) amplitude pairs and
therefore has a computational complexity of
\(\mathcal{O}(B^2)\) with respect to the colour-basis size. The folded
triangular formulation evaluates \(B(B+1)/2\) matrix positions, reducing the
number of explicitly traversed pairs while preserving the quadratic
asymptotic complexity.

For the benchmark processes considered here, the colour bases contain 6, 24,
and 120 amplitudes for the one-, two-, and three-jet kernels, respectively.
The increase in basis size therefore produces a substantial increase in both
the arithmetic workload and the amount of coefficient and amplitude data that
must be accessed by the kernel.

For the largest kernel, the implementation is organized as a multi-stage
dataflow architecture. Separate modules perform input loading, data unpacking,
colour-matrix contraction, reduction, and output packing, allowing these
stages to operate concurrently on different data items. The colour matrix is
stored on chip as a constant ROM array.

Within the matrix-product stage, loop unrolling is used to evaluate several
iterations of the colour contraction in parallel. Instead of reusing a single
multiply--accumulate datapath for every matrix position, the unrolled loop
exposes multiple colour terms to concurrent arithmetic operators. This
increases throughput at the cost of additional arithmetic resources and
routing complexity.

The amplitude arrays are partially partitioned to provide the memory bandwidth
required by these parallel operators. Without partitioning, several unrolled
iterations could attempt to access the same array storage in the same clock
cycle, limiting the achievable parallelism. Partial partitioning distributes
the array elements across multiple independently accessible storage banks,
allowing several amplitudes to be read concurrently without fully replicating
the complete array.

Loop unrolling and array partitioning are therefore applied together: the
unrolling determines how many colour terms can be processed concurrently,
while the partitioning provides the corresponding parallel data accesses. The
degree of parallelism remains limited by the available DSP and logic
resources, the number of memory ports, routing congestion, and the timing
target. This trade-off is particularly relevant for the three-jet kernel,
whose 120-amplitude colour basis is considerably larger than those of the
one- and two-jet benchmarks.

This organization increases the available arithmetic parallelism while
keeping the implementation within the resources of the target FPGA.
Accordingly, the reported performance should be interpreted as the performance
of the isolated colour-reduction kernel rather than as an end-to-end
matrix-element or event-generation speedup. The colour kernel therefore
provides a controlled target for studying selective FPGA acceleration without
requiring the complete matrix-element pipeline to be mapped onto the device.

\subsubsection{Numerical representation}

The FPGA implementations considered in this work employ different numerical
representations depending on the complexity and structure of the target
kernel. In all cases, the numerical format is chosen as a compromise between
hardware cost and numerical accuracy, with the goal of preserving sufficient
precision while keeping resource usage within the limits of the target
device. The formats were selected through an iterative, manually assisted
procedure based on numerical comparisons with the corresponding CPU reference
and subsequent HLS synthesis. No automatic word-length optimiser was used in
this work.

For the full matrix-element implementation of
\(e^+e^- \to \mu^+\mu^-\), a fixed-point numerical representation is adopted
throughout the main computational pipeline. This choice is motivated by the
reduced hardware cost of fixed-point arithmetic compared to floating-point
arithmetic, while the resulting numerical accuracy for the benchmark process
is evaluated in Section~\ref{sec:numerical_validation_full}. Multiple
fixed-point formats are employed across the different stages of the pipeline:
a 24-bit fixed-point format is used for both phase-space generation and the
internal matrix-element computation, while wider formats are used at the
memory interface and for the accumulation over helicity configurations. This
organization reflects the fact that the scaled kinematic variables and
internal amplitudes can be represented within a compact range, whereas
accumulation and output stages require additional dynamic-range headroom to
accommodate growth in magnitude.

For the colour-algebra kernels associated with
\(gg \to t\bar{t}+X\) processes, the numerical representation is adapted to
the complexity of each configuration. Unless otherwise stated,
floating-point arithmetic refers to single-precision FP32 arithmetic. FP32 is
used for the one- and two-jet cases, where the colour bases contain 6 and 24
amplitudes, respectively, and the corresponding kernels can be implemented
without excessive resource pressure. For the three-jet case, where the colour
basis increases to 120 amplitudes, a floating-point implementation becomes
significantly more demanding in terms of FPGA resources. To address this, a
fixed-point representation with explicit scaling is adopted, using compact
formats for the input amplitudes and stage-specific formats for the
colour-matrix coefficients, reduced intermediate values, and accumulation.
This reflects the different numerical requirements of each stage: the input
amplitudes can be stored in a reduced range, whereas the matrix--vector
products and final reduction require additional integer width to provide
sufficient dynamic-range headroom.

The selected formats therefore represent an empirical compromise between
synthesis feasibility, dynamic-range requirements, and the numerical behaviour
observed for the considered benchmarks. More specifically, the design
exploration considered four main factors: the dynamic range of the
corresponding physical quantities, the numerical growth expected in
intermediate products and accumulations, the numerical agreement with the CPU
reference, and the resource and timing constraints of the target FPGA.
Compact formats were preferred whenever possible in order to limit
arithmetic-resource usage and routing pressure, while wider types were used
for accumulation, rescaling, and matrix--vector product stages, where greater
numerical growth was expected.

The notation \texttt{ap\_fixed<W,I>} denotes a signed fixed-point type with
\(W\) total bits and \(I\) integer bits, including the sign bit. The number of
fractional bits is therefore
\begin{equation}
F = W-I,
\end{equation}
giving a quantisation step
\begin{equation}
\Delta = 2^{-F}
\end{equation}
and a theoretical representable range
\begin{equation}
\left[-2^{I-1},\,2^{I-1}-2^{-F}\right].
\end{equation}
The integer width determines the available dynamic range, whereas the
fractional width determines the numerical resolution. The retained
fixed-point types use truncation and wrap-around semantics,
\texttt{AP\_TRN} and \texttt{AP\_WRAP}, either explicitly or through the
Vitis HLS defaults. Consequently, assignments to narrower fixed-point types
discard the least-significant fractional bits rather than applying rounding,
and no explicit saturation mode is used.

The main numerical formats used in the fixed-point implementations are
summarized in Table~\ref{tab:fixed_point_formats}.

\begin{table*}[t]
\centering
\caption{Fixed-point numerical formats used in the FPGA implementations.
For \texttt{ap\_fixed<W,I>}, \(F=W-I\) and the quantisation step is
\(2^{-F}\).}
\label{tab:fixed_point_formats}
\resizebox{\textwidth}{!}{
\begin{tabular}{llllll}
\toprule
\textbf{Implementation}
& \textbf{Quantity / stage}
& \textbf{Data type}
& \textbf{\(F\)}
& \textbf{Quantisation step}
& \textbf{Role} \\
\midrule

Full ME \(e^+e^- \to \mu^+\mu^-\)
& Memory interface and output data
& \texttt{ap\_fixed<32,14>}
& 18
& \(2^{-18}\)
& External memory interface and output storage \\

Full ME \(e^+e^- \to \mu^+\mu^-\)
& Phase-space generation
& \texttt{ap\_fixed<24,8>}
& 16
& \(2^{-16}\)
& RAMBO-based momentum construction \\

Full ME \(e^+e^- \to \mu^+\mu^-\)
& Matrix-element computation
& \texttt{ap\_fixed<24,8>}
& 16
& \(2^{-16}\)
& Wavefunctions, propagators, and amplitudes \\

Full ME \(e^+e^- \to \mu^+\mu^-\)
& Helicity accumulation
& \texttt{ap\_fixed<32,12>}
& 20
& \(2^{-20}\)
& Summation over helicity contributions \\

Full ME \(e^+e^- \to \mu^+\mu^-\)
& Momentum output rescaling
& \texttt{ap\_fixed<48,24>}
& 24
& \(2^{-24}\)
& Temporary wide type used during output rescaling \\

\midrule

Colour \(gg \to t\bar{t}+3j\)
& Input amplitude components
& \texttt{ap\_fixed<16,4>}
& 12
& \(2^{-12}\)
& Real and imaginary parts of colour-flow amplitudes \\

Colour \(gg \to t\bar{t}+3j\)
& Colour-matrix coefficients
& \texttt{ap\_fixed<24,7>}
& 17
& \(2^{-17}\)
& Entries of the colour matrix \\

Colour \(gg \to t\bar{t}+3j\)
& Reduced intermediate values
& \texttt{ap\_fixed<22,10>}
& 12
& \(2^{-12}\)
& Intermediate matrix--vector results \\

Colour \(gg \to t\bar{t}+3j\)
& Accumulation
& \texttt{ap\_fixed<28,15>}
& 13
& \(2^{-13}\)
& Colour-index reduction \\

Colour \(gg \to t\bar{t}+3j\)
& Output representation
& \texttt{ap\_int<32>} / \texttt{ap\_uint<128>}
& --
& Integer representation
& Raw output value and packed output word \\

\bottomrule
\end{tabular}
}
\end{table*}

A key aspect of the full matrix-element implementation is the explicit scaling
of particle momenta and mass parameters. The four-momenta are scaled by a
constant factor chosen as a power of two,
\begin{equation}
p_{\mathrm{scaled}} = \frac{p}{S},
\qquad
S = 2^{10},
\label{eq:momentum_scaling}
\end{equation}
which reduces the dynamic range of the kinematic variables while retaining
their relative scale, subject to fixed-point quantisation. In the HLS
implementation, the dimensional inputs are supplied directly in pre-scaled
form, so Eq.~\eqref{eq:momentum_scaling} does not introduce a run-time
division. At the output stage, the original momentum scale is restored by
multiplication by \(2^{10}\), expressed in C++ using a wider temporary
fixed-point type. Vitis HLS reduces this constant multiplication in the
generated RTL to bit selection and the concatenation of ten zero
least-significant bits, without requiring a general-purpose divider or
multiplier.

The full matrix-element computation is performed using complex fixed-point
arithmetic, in which both real and imaginary components are explicitly
represented. The squared matrix element is then expressed in terms of
real-valued combinations of these components, allowing the final reduction to
be implemented using real arithmetic. In addition, a wider fixed-point
representation is used at the memory-interface level in order to support the external data representation and the subsequent
output-rescaling operation.

For the full matrix-element implementation, no single global error threshold
was used to select all internal formats. Candidate configurations were
evaluated according to their numerical agreement with the software reference
and their synthesis feasibility. The final implementation was subsequently
validated over the complete benchmark dataset reported in
Section~\ref{sec:numerical_validation_full}.

For the fixed-point colour-algebra implementation, the input colour-flow
amplitudes are represented using compact 16-bit fixed-point real and imaginary
components. Before their conversion to fixed point, the input components are
normalised using
\begin{equation}
S_{\mathrm{in}} = 8.6 \times 10^{-8}.
\end{equation}
The corresponding physical output is recovered using
\begin{equation}
S_{\mathrm{out}} = S_{\mathrm{in}}^2.
\end{equation}
These scaling operations are performed outside the computational kernel,
which operates directly on the normalised fixed-point representations.

The colour-matrix coefficients, reduced intermediate values, and final colour
reduction use dedicated stage-specific fixed-point formats. This
mixed-precision organization reduces the hardware pressure associated with
the substantially larger three-jet colour basis while retaining the numerical
range and resolution required by the different stages of the contraction.

There is no single dynamic truncation threshold applied throughout the
design. Precision reduction occurs deterministically whenever a value is
assigned to a type with fewer fractional bits, according to the
\texttt{AP\_TRN} quantisation mode. In the principal narrowing operations of
the three-jet kernel, the coefficient--amplitude product is reduced from
29 to 13 fractional bits, the reduced row value from 13 to 12 fractional
bits, and the subsequent amplitude--row product from 24 to 13 fractional
bits. These conversions therefore discard 16, 1, and 11 fractional bits,
respectively. The retained configurations use \texttt{AP\_WRAP} rather than
saturation, and the integer widths and scaling factors were selected to
provide dynamic-range headroom for the validated workloads.

During the development of the three-jet kernel, narrower 8-bit and 12-bit
input-amplitude representations were also evaluated, but they did not satisfy
the numerical regression criteria used during design exploration. The retained 16-bit representation satisfied the \(1\%\) relative-error
criterion used in the archived one-event, four-output development regression
test. This test was used to guide format selection, whereas broader numerical
validation is reported in
Section~\ref{sec:numerical_validation_colour}.

The precision--hardware trade-off was managed iteratively. Candidate formats
were first evaluated through software simulation against the CPU reference,
and the retained configurations were subsequently assessed through HLS
synthesis in terms of resource usage, pipeline performance, and timing
feasibility. The reported formats should therefore be understood as validated
empirical design choices rather than as the result of a formal global
optimisation.

Although this process could be partially automated through dynamic-range
profiling, parameterised word-length sweeps, numerical acceptance thresholds,
and automatic extraction of HLS results, no fully automatic
format-selection or Pareto-optimisation framework was used in this work.

This adaptive strategy makes it possible to match the numerical representation
to the complexity of the target workload: FP32 arithmetic provides a
straightforward implementation with good numerical behaviour for the simpler
kernels, whereas fixed-point arithmetic enables a more resource-efficient
implementation for the most demanding case. The numerical impact of these
choices is evaluated in Sections~\ref{sec:numerical_validation_full} and
\ref{sec:numerical_validation_colour}.

\subsection{Benchmark setup and evaluation metrics}
\label{sec:benchmark_setup}
The FPGA, CPU, and GPU implementations are evaluated in terms of both computational performance and energy efficiency. The primary performance metric is throughput, defined as the number of events processed per unit time. In addition, power consumption is measured in order to derive the energy per event, which provides a normalized metric for comparing different hardware architectures.

The FPGA execution is managed through the Xilinx Runtime (XRT), with kernels invoked from a Python-based host application~\cite{AMD:XRTDocumentation}. Execution times are measured using high-resolution wall-clock timers on the host side, while FPGA kernel activity is monitored through the runtime environment.

Power measurements for the FPGA are obtained using the monitoring capabilities provided by the platform through XRT. For the CPU implementation, power consumption is measured using the RAPL interface~\cite{Intel:RAPL}, while GPU power measurements are collected using the \texttt{nvidia-smi} utility~\cite{NVIDIA:nvidia-smi}. Although these measurements rely on platform-specific monitoring interfaces, they are used consistently to estimate device-level average power under sustained workload conditions.

The FPGA results are compared against CPU and GPU implementations available within the \mgfive\ framework. Within each benchmark scenario, all measurements are performed using identical input datasets and physics configurations to ensure a fair and consistent comparison across platforms. Unless otherwise stated, initialization overheads and host--device data-transfer times are excluded from the reported performance metrics in order to isolate the computational behavior of the kernels themselves. All measurements are averaged over multiple runs in order to reduce statistical fluctuations.

\section{Results}

\subsection{Full matrix-element implementation for \(e^+e^- \to \mu^+\mu^-\)}

\subsubsection{Numerical validation}
\label{sec:numerical_validation_full}

We first assess the numerical accuracy of the full FPGA implementation against reference results obtained with \mgfive. The validation includes both phase-space generation and matrix-element evaluation. The reported statistics are based on a sample of \(3\times 10^7\) randomly generated events. The CPU reference is evaluated in double precision, whereas the FPGA implementation uses reduced-precision fixed-point representations. In addition to relative errors, absolute error metrics are also reported, since they are particularly informative in regions where the reference value becomes small and relative differences may become less representative. To provide further context for these absolute deviations, we also consider the least significant bit (LSB) of the main fixed-point formats used in the implementation. For a type \texttt{ap\_fixed<W,I>}, the LSB is given by \(2^{-(W-I)}\), and represents the nominal resolution of the corresponding quantized quantity. In the present implementation, however, the reported absolute deviations reflect the cumulative effect of quantization across multiple pipeline stages, including input representation, intermediate products, casts, accumulation, and final reconstruction. They should therefore not be interpreted as deviations arising from a single quantization step or from the LSB of a single format alone.

For the full FPGA implementation of the \(e^+e^- \to \mu^+\mu^-\) process, two main sources of numerical discrepancy can be identified: the precision of the generated momenta and the precision of the matrix-element evaluation itself.

The momenta are generated on the FPGA using a fixed-point representation within the RAMBO algorithm. Their comparison with the CPU reference shows a mean relative error of 0.020\%, with a standard deviation of 0.027\% and a maximum observed relative deviation of 0.138\%. This indicates that the kinematic configuration is reproduced with high accuracy.

The matrix-element evaluation exhibits a mean relative error of 0.160\%, with a standard deviation of 0.256\% and a maximum observed relative deviation of 1.352\%, as summarized in Table~\ref{tab:numerical_validation_full}, which reports both relative and absolute error metrics. Despite the comparison against a double-precision CPU reference, the observed discrepancies remain small, suggesting that, for this benchmark, the reduced-precision FPGA implementation remains in close numerical agreement with the full matrix-element calculation.

% For reference, the dominant internal fixed-point format used in the full matrix-element pipeline is \texttt{ap\_fixed<24,8>}, which corresponds to an LSB of \(2^{-16} \approx 1.53\times 10^{-5}\) in scaled units. This value provides a reference for the local numerical resolution of the dominant internal representation used in the pipeline. The absolute errors reported in Table~\ref{tab:numerical_validation_full} should therefore be interpreted as the cumulative effect of quantization across the full fixed-point pipeline, rather than as deviations arising from a single quantization step.

\begin{table*}[t]
\centering
\caption{Numerical validation of the full matrix-element implementation for \(e^+e^- \to \mu^+\mu^-\), using \(3\times 10^7\) events.}
\label{tab:numerical_validation_full}
\resizebox{\textwidth}{!}{
\begin{tabular}{lcccccc}
\toprule
\textbf{Quantity} &
\makecell{\textbf{Mean rel.}\\\textbf{error [\%]}} &
\makecell{\textbf{Std. rel.}\\\textbf{error [\%]}} &
\makecell{\textbf{Max rel.}\\\textbf{error [\%]}} &
\makecell{\textbf{Mean abs.}\\\textbf{error}} &
\makecell{\textbf{Std. abs.}\\\textbf{error}} &
\makecell{\textbf{Max abs.}\\\textbf{error}} \\
\midrule
Momenta & 0.020 & 0.027 & 0.138 & $5.47 \times 10^{-2}$ & $1.09 \times 10^{-1}$ & $5.11$ \\
Matrix-element & 0.160 & 0.256 & 1.352 & $2.72 \times 10^{-5}$ & $6.35 \times 10^{-5}$ & $4.54 \times 10^{-4}$ \\
\bottomrule
\end{tabular}
}
\end{table*}

\subsubsection{Performance evaluation}
\label{sec:performance_full}

The performance of the proposed full matrix-element implementation is evaluated in terms of measured execution time and effective throughput. The reported execution times correspond to the mean over 10 independent runs for a sample of \(3\times 10^7\) events, and the associated variability is quantified through the standard deviation. Host-side initialization, data transfers, and other platform-specific overheads are excluded, so that the reported values reflect the pure computational performance of the evaluated implementations.

\begin{figure*}[t]
    \centering
    \includegraphics[width=\textwidth]{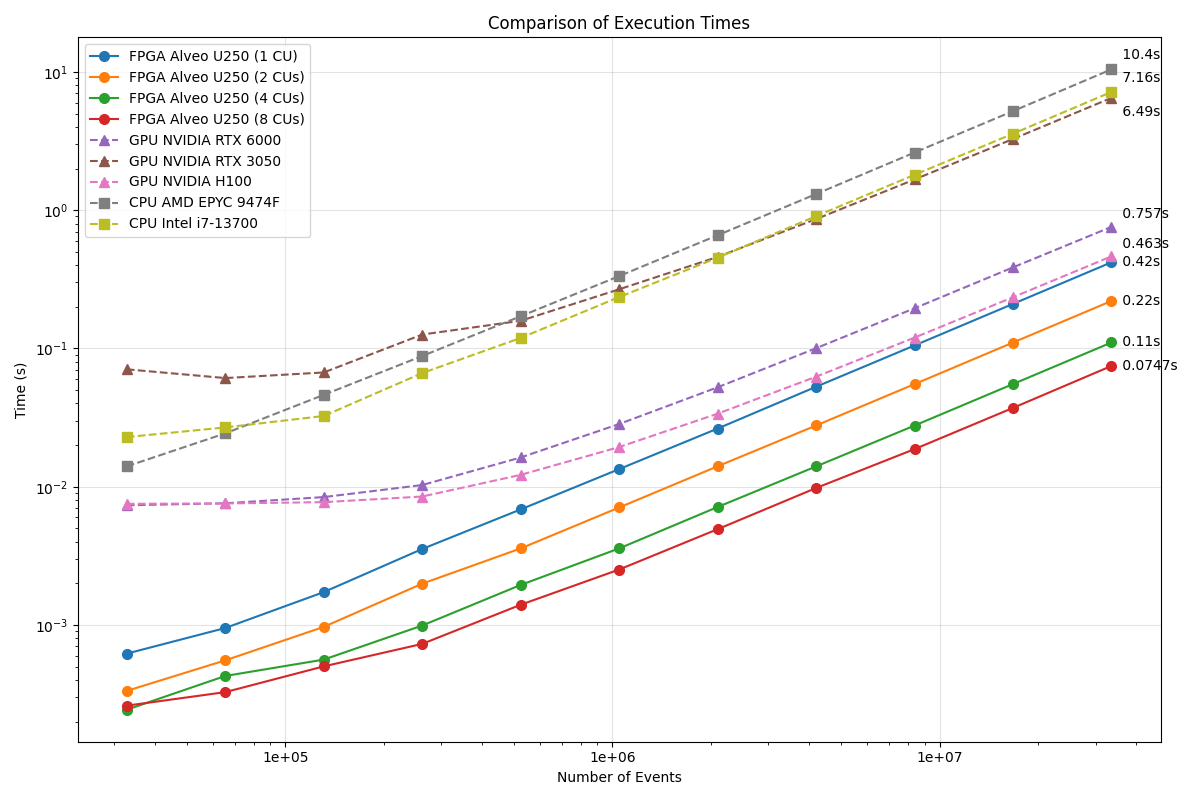}
    \caption{Execution time as a function of the number of processed events for CPU, GPU, and FPGA implementations. FPGA results are shown for different CU configurations.}
    \label{fig:execution_time}
\end{figure*}

Figure~\ref{fig:execution_time} shows the execution time as a function of the number of processed events for the \(e^+e^- \to \mu^+\mu^-\) process, comparing CPU, GPU, and FPGA implementations. The CPU results correspond to the AMD EPYC 9474F and Intel i7-13700 processors, while the GPU results are obtained using NVIDIA RTX 3050 8GB, RTX 6000, and H100 devices. The FPGA implementation is evaluated on an AMD Alveo U250 accelerator.

The CPU implementations were executed using 24 software threads. The Intel
Core i7-13700 contains 16 physical cores, comprising 8 performance cores and
8 efficiency cores, and exposes 24 hardware threads. Therefore, the
24-thread execution uses all logical processors available on this platform.
The AMD EPYC 9474F contains 48 physical cores and exposes 96 hardware threads;
the reported EPYC measurement therefore corresponds to a fixed 24-thread
execution rather than to full-socket utilisation. In both cases, the
generated implementation exploits event-level parallelism by evaluating
independent phase-space points concurrently.

The generated CUDA implementation was launched using 256 blocks with
256 threads per block on each evaluated GPU. Independent phase-space points
are processed concurrently across the CUDA thread grid. This launch
configuration was kept fixed for the RTX 3050, RTX 6000, and H100
measurements.

The results show distinct scaling trends across the considered architectures. The CPU implementations exhibit the largest execution times and an approximately linear dependence on the number of processed events. The GPU implementations achieve substantially lower execution times, although a noticeable non-scaling component is observed for small event samples.

The FPGA implementation shows a highly regular scaling trend, with execution
time increasing approximately linearly with the number of processed events
and comparatively small fixed overheads. This behaviour is consistent with
the streaming and deeply pipelined organisation of the design. Multiple FPGA
configurations with different numbers of compute units (CUs) are evaluated.
Increasing the number of CUs produces close-to-linear scaling up to four CUs,
whereas the 8-CU configuration exhibits a more pronounced sublinear trend.
Nevertheless, the 8-CU configuration provides the highest measured
throughput and outperforms all evaluated CPU and GPU baselines for the
considered event samples.

Table~\ref{tab:throughput} summarizes the effective throughput measured for the \(3\times 10^7\)-event sample, together with the mean execution time over 10 runs and its standard deviation, providing a direct comparison across platforms.

\begin{table*}[t]
\centering
\caption{Effective throughput for the full matrix-element computation
(\(3\times 10^7\) events). Reported execution times correspond to the mean
over 10 runs. The speedup of the 8-CU FPGA configuration is calculated as
\(S_p=\overline{T}_p/\overline{T}_{\mathrm{FPGA,8CU}}\), where values greater
than one indicate that the 8-CU FPGA implementation is faster than the
corresponding platform.}
\label{tab:throughput}
\resizebox{\textwidth}{!}{
\begin{tabular}{lcccc}
\toprule
\textbf{Platform}
& \textbf{Average time [s]}
& \textbf{Std. [s]}
& \textbf{Throughput [events/s]}
& \textbf{Speedup of 8-CU FPGA} \\
\midrule
CPU AMD EPYC 9474F
& 10.394 & 0.016 & $2.89 \times 10^6$ & $138.59\times$ \\

CPU Intel i7-13700
& 7.165 & 0.030 & $4.19 \times 10^6$ & $95.53\times$ \\

GPU NVIDIA RTX 3050 8GB
& 6.491 & 0.050 & $4.62 \times 10^6$ & $86.55\times$ \\

GPU NVIDIA RTX 6000
& 0.757 & 0.004 & $3.96 \times 10^7$ & $10.09\times$ \\

GPU NVIDIA H100
& 0.476 & 0.001 & $6.31 \times 10^7$ & $6.35\times$ \\

FPGA Alveo U250 (1 CU)
& 0.422 & $<$ 0.001 & $7.15 \times 10^7$ & $5.63\times$ \\

FPGA Alveo U250 (2 CUs)
& 0.224 & $<$ 0.001 & $1.36 \times 10^8$ & $2.99\times$ \\

FPGA Alveo U250 (4 CUs)
& 0.113 & $<$ 0.001 & $2.65 \times 10^8$ & $1.51\times$ \\

FPGA Alveo U250 (8 CUs)
& 0.075 & $<$ 0.001 & $4.01 \times 10^8$ & $1.00\times$ \\
\bottomrule
\end{tabular}
}
\end{table*}

The throughput values are consistent with the execution-time trends observed
in Figure~\ref{fig:execution_time}. The 8-CU FPGA configuration reaches
\(4.01\times10^8\) events/s and achieves speedups of approximately
\(138.6\times\) and \(95.5\times\) relative to the measured AMD EPYC 9474F
and Intel i7-13700 baselines, respectively. Relative to the evaluated GPU
implementations, the speedup ranges from \(6.35\times\) over the NVIDIA H100
to \(86.55\times\) over the NVIDIA RTX 3050.

\subsubsection{Throughput efficiency analysis}
\label{sec:throughput_efficiency_full}

To better interpret the measured performance, the experimental results are compared with the ideal throughput expected from the target clock frequency and the pipeline initiation interval (II). For a design operating at a clock frequency \(f_{\mathrm{clk}}\), the ideal throughput of a single compute unit is given by
\begin{equation}
\mathrm{Throughput}_{\mathrm{ideal,\,CU}} = \frac{f_{\mathrm{clk}}}{\mathrm{II}},
\end{equation}
so that the corresponding ideal aggregate throughput for a design with \(N_{\mathrm{CU}}\) compute units is
\begin{equation}
\mathrm{Throughput}_{\mathrm{ideal}} = \frac{N_{\mathrm{CU}} \, f_{\mathrm{clk}}}{\mathrm{II}}.
\end{equation}
The corresponding ideal time per event is therefore
\begin{equation}
t_{\mathrm{ideal}} = \frac{\mathrm{II}}{N_{\mathrm{CU}} \, f_{\mathrm{clk}}}.
\end{equation}
The efficiency of the measured implementation with respect to this ideal limit is then defined as
\begin{equation}
\eta = \frac{\mathrm{Throughput}_{\mathrm{measured}}}{\mathrm{Throughput}_{\mathrm{ideal}}}.
\end{equation}
These expressions refer to the ideal steady-state behavior of the pipeline and therefore neglect startup, drain, and synchronization overheads.

To complement the throughput analysis, Table~\ref{tab:fpga_impl_summary} summarizes the main post-implementation timing characteristics of the evaluated FPGA kernels, including target and achieved clock frequencies, achieved initiation interval, pipeline latency, and worst negative slack.

\begin{table}[t]
\centering
\caption{Post-implementation timing summary of the evaluated FPGA kernels.}
\label{tab:fpga_impl_summary}
\resizebox{\columnwidth}{!}{%
\begin{tabular}{lccccc}
\toprule
\textbf{Process} &
\makecell{\textbf{Target clk}\\\textbf{[MHz]}} &
\makecell{\textbf{Achieved clk}\\\textbf{[MHz]}} &
\makecell{\textbf{Achieved}\\\textbf{II}} &
\makecell{\textbf{Latency}\\\textbf{[cycles]}} &
\makecell{\textbf{WNS}\\\textbf{[ns]}} \\
\midrule
$gg \to t\bar{t}g$              & 100 & 100.9  & 1 & 182 & 0.097 \\
$gg \to t\bar{t}gg$             & 100 & 102.2 & 1 & 290 & 1.141 \\
$gg \to t\bar{t}ggg$            & 100 & 100.0 & 4 & 157 & 0.367 \\
$e^+e^- \to \mu^+\mu^-$ (1 CU)  & 100 & 110.1 & 1 & 141 & 0.567 \\
$e^+e^- \to \mu^+\mu^-$ (8 CUs) & 100 & 110.1 & 1 & 141 & 0.057 \\
\bottomrule
\end{tabular}%
}
\end{table}

These results confirm that the full \(e^+e^- \to \mu^+\mu^-\) implementation sustains \(\mathrm{II}=1\) for both the 1-CU and 8-CU configurations at an achieved operating frequency above 100~MHz.

For the full matrix-element implementation, the target operating frequency is 100~MHz. Under ideal steady-state conditions, a design achieving \(\mathrm{II}=1\) would process one event per clock cycle, corresponding to an ideal throughput of \(10^8\) events/s per compute unit, or equivalently one event every 10~ns for each compute unit.

Table~\ref{tab:full_ideal} compares the measured throughput of the different FPGA configurations with the corresponding ideal aggregate limits. The single-CU implementation reaches about 71.5\% of the theoretical maximum, while the 2-CU and 4-CU configurations remain close to 68\%. For the 8-CU case, the aggregate throughput reaches \(4.01 \times 10^8\) events/s, corresponding to 50.1\% of the ideal limit of \(8.00 \times 10^8\) events/s.

\begin{table*}[t]
\centering
\caption{Measured throughput of the full matrix-element implementation compared with the ideal throughput at 100~MHz and \(\mathrm{II}=1\).}
\label{tab:full_ideal}
\resizebox{\textwidth}{!}{%
\begin{tabular}{lcccc}
\toprule
\textbf{Configuration} &
\makecell{\textbf{Measured Throughput}\\\textbf{[events/s]}} &
\makecell{\textbf{Ideal Throughput}\\\textbf{[events/s]}} &
\makecell{\textbf{Efficiency}\\\textbf{[\%]}} &
\makecell{\textbf{Ideal time/evt}\\\textbf{[ns]}} \\
\midrule
FPGA Alveo U250 (1 CU) & $7.15 \times 10^7$ & $1.00 \times 10^8$ & 71.5 & 10 \\
FPGA Alveo U250 (2 CUs) & $1.36 \times 10^8$ & $2.00 \times 10^8$ & 68.0 & 5 \\
FPGA Alveo U250 (4 CUs) & $2.65 \times 10^8$ & $4.00 \times 10^8$ & 66.3 & 2.5 \\
FPGA Alveo U250 (8 CUs) & $4.01 \times 10^8$ & $8.00 \times 10^8$ & 50.1 & 1.25 \\
\bottomrule
\end{tabular}%
}
\end{table*}

These results show that the implementation remains relatively close to the ideal scaling trend for the smaller CU configurations, but departs from it more visibly as the number of compute units increases. This behavior is expected, since the ideal model assumes perfect overlap between pipeline stages, no stalls, and continuous data availability. In practice, the effective throughput is limited by the slowest stage of the dataflow pipeline, as well as by memory-access overheads, data movement, routing pressure, and scheduling constraints in the matrix-element kernel.

\subsubsection{Power consumption and energy efficiency}

For the full matrix-element benchmark, device-level power was recorded under
sustained workload conditions using the measurement methodology described in
Section~\ref{sec:benchmark_setup}. The reported values correspond to the CPU
package, GPU card, and FPGA accelerator, respectively, and exclude the power
consumption of the host system. Average power was calculated over the
steady-state execution intervals used for the corresponding performance
measurements.

To evaluate energy efficiency, we consider the energy required to process a single event, defined as
\begin{equation}
E_{\mathrm{event}} = \frac{P_{\mathrm{avg}}}{\mathrm{Throughput}},
\end{equation}
where \(P_{\mathrm{avg}}\) is the average power consumption during execution and the throughput is expressed in events per second.

\begin{figure*}[t]
    \centering
    \includegraphics[width=\textwidth]{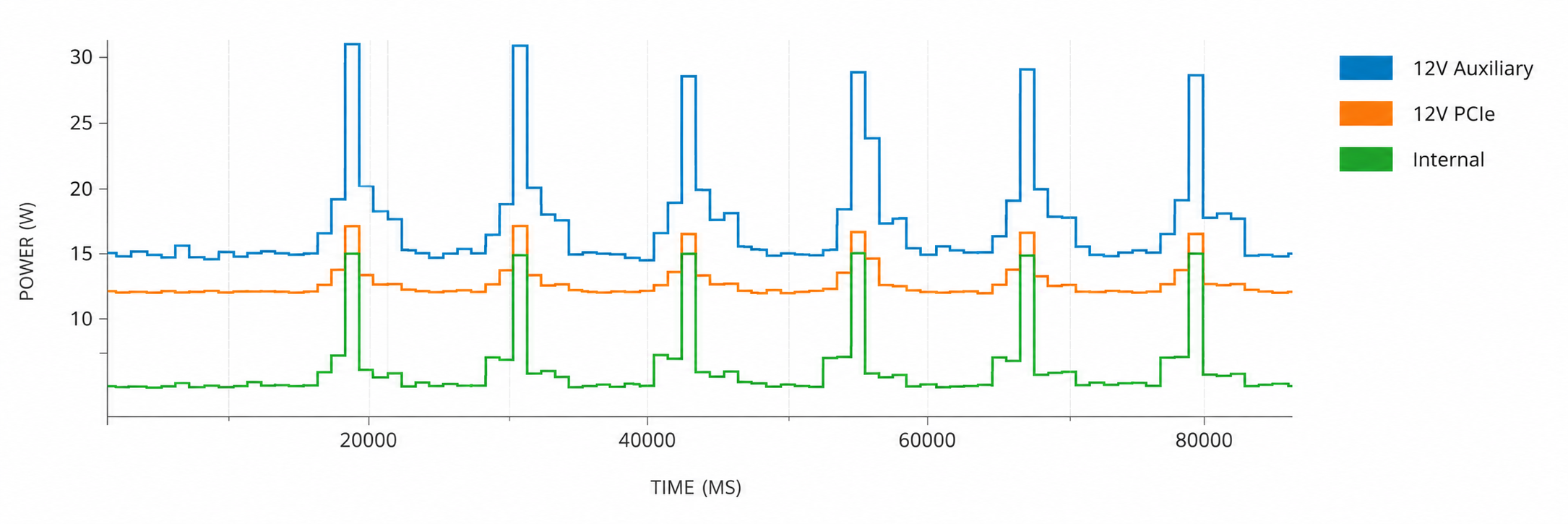}
    \caption{Measured FPGA power trace for the full matrix-element implementation during the processing of 1.8 billion events, executed in batches of 300 million events. The three reported components correspond to the 12V auxiliary rail, the 12V PCIe rail, and the internal power contribution.}
    \label{fig:fpga_power_trace}
\end{figure*}

Figure~\ref{fig:fpga_power_trace} illustrates the temporal evolution of the FPGA power during the processing of 1.8 billion events, executed in six consecutive batches of 300 million events each. The repeated peaks reflect the regular execution pattern of the accelerator across batches, while the overall stability of the trace indicates a sustained steady-state operating regime during the measurement window.

For the CPU platform, the Intel i7-13700 is used as the reference system for the reported power measurements. Table~\ref{tab:energy_efficiency} summarizes the measured average power, throughput, and resulting energy per event for the evaluated platforms.

\begin{table}[t]
\centering
\caption{Power consumption and energy per event for the full matrix-element implementation.}
\label{tab:energy_efficiency}
\resizebox{\columnwidth}{!}{
\begin{tabular}{lccc}
\toprule
\textbf{Platform} & 
\makecell{\textbf{Average power}\\\textbf{[W]}} &
\makecell{\textbf{Throughput}\\\textbf{[events/s]}} &
\makecell{\textbf{Average Energy/event}\\\textbf{[$\mu$J]}} \\
\midrule
CPU (Intel i7-13700) & 110 & $4.19 \times 10^6$ & 26.3 \\
GPU (RTX 3050) & 46 & $4.62 \times 10^6$ & 9.96 \\
GPU (RTX 6000) & 88.6 & $3.96 \times 10^7$ & 2.23 \\
GPU (H100) & 92.27 & $6.31 \times 10^7$ & 1.46 \\
FPGA (Alveo U250, 8 CUs) & 72 & $4.01 \times 10^8$ & 0.18 \\
\bottomrule
\end{tabular}
}
\end{table}

The results show substantial differences in energy efficiency across the evaluated platforms. The CPU implementation exhibits the largest energy cost per event, while the GPU implementations achieve a clear improvement but remain in the \(\mu\)J/event range. The FPGA implementation achieves the lowest energy per event among the
evaluated platforms, reaching \(0.18\,\mu\mathrm{J}\) per event in the
8-CU configuration, compared with \(2.23\,\mu\mathrm{J}\),
\(1.46\,\mu\mathrm{J}\), and \(9.96\,\mu\mathrm{J}\) for the RTX 6000,
H100, and RTX 3050, respectively, and \(26.3\,\mu\mathrm{J}\) for the CPU
baseline.

This energy-efficiency advantage is consistent with the streaming dataflow organization of the FPGA design, which enables sustained high-throughput execution with limited control overhead, and with the use of fixed-point arithmetic in the full matrix-element pipeline.

Overall, under the stated compute-only measurement methodology, the evaluated
FPGA implementation combines high throughput with substantially lower
device-level energy per event than the considered CPU and GPU baselines.
These results demonstrate the potential of FPGA acceleration for the
matrix-element benchmark studied here, while an integrated assessment
including host-side execution, data transfers, and CPU--FPGA orchestration
would be required to determine the energy benefit at the level of a complete
event-generation workflow.

\subsubsection{Resource utilization and scalability}
\label{sec:resources_scalability_full}

Table~\ref{tab:resource_full} summarizes the FPGA resource utilization for the full matrix-element implementation in the 1-CU and 8-CU configurations. These two cases represent the baseline design and the most aggressive parallel configuration considered in this work. The reported values correspond to absolute resource usage on the target device, together with the corresponding fraction of the total available resources.

\begin{table*}[t]
\centering
\caption{FPGA resource utilization for the full matrix-element implementation.}
\label{tab:resource_full}
\resizebox{\textwidth}{!}{
\begin{tabular}{lcccc}
\toprule
\textbf{Configuration} & \textbf{LUTs} & \textbf{FFs} & \textbf{DSPs} & \textbf{BRAM} \\
\midrule
FPGA Alveo U250 (1 CU) & 175189 (9.95\%) & 350589 (9.58\%) & 1129 (9.09\%) & 31 (0.56\%) \\
FPGA Alveo U250 (8 CUs) & 641552 (36.46\%) & 700871 (19.15\%) & 8656 (69.67\%) & 128 (2.35\%) \\
\bottomrule
\end{tabular}
}
\end{table*}

The increase in resource utilization from 1 CU to 8 CUs reflects the replication of the matrix-element kernel across multiple units. Among all resources, DSP usage shows the strongest scaling with the number of instantiated kernels, reaching nearly 70\% of the available device resources in the 8-CU configuration. This indicates that arithmetic operations dominate the design and that DSP availability is the primary factor limiting further replication.

LUT and FF utilization also increase significantly, due not only to the replication of the compute logic itself but also to the additional control and dataflow infrastructure required to sustain parallel execution. In contrast, BRAM usage remains comparatively modest in the full matrix-element design.

These trends are consistent with the performance results. Although throughput improves substantially as the number of compute units increases, the scaling departs from the ideal linear regime for the largest configuration. This behavior can be attributed to the growing impact of DSP saturation, routing congestion, timing-closure pressure, and memory-access overheads as the design approaches the practical limits of the device.

\subsection{Colour-algebra kernels for \(gg \to t\bar{t}+X\)}

\subsubsection{Numerical validation}
\label{sec:numerical_validation_colour}

For the colour-algebra kernels associated with \(gg \to t\bar{t}+X\) processes, the numerical accuracy is evaluated independently for different jet multiplicities. The reported statistics are based on a sample of \(3\times 10^7\) randomly generated events. The CPU reference is evaluated in double precision, whereas the FPGA implementations use reduced-precision numerical representations. In addition to relative errors, absolute error metrics are also reported, since they are particularly informative in regions where the reference value becomes small and relative differences may become less representative.

The results are summarized in Table~\ref{tab:numerical_validation_colour}, which reports mean, standard deviation, and maximum values for both relative and absolute error metrics.

The floating-point implementations for the one- and two-jet cases reproduce the reference results with very small numerical differences. For \(t\bar{t}+1\) jet, the mean relative error is $5.432\times 10^{-5}\%$, with a standard deviation of $9.16\times 10^{-5}\%$ and a maximum observed deviation of $6.678\times 10^{-4}\%$. For \(t\bar{t}+2\) jets, the mean relative error is $8.573\times 10^{-4}\%$, with a standard deviation of $1.65\times 10^{-3}\%$ and a maximum observed deviation of $6.60\times 10^{-2}\%$. These results indicate that, when both the FPGA and CPU implementations use floating-point arithmetic, the FPGA reproduces the reference calculation with very close numerical agreement.

For the three-jet case, the FPGA kernel uses an \texttt{ap\_fixed} representation with explicit scaling, whereas the CPU reference is evaluated in double precision. This choice is motivated by the substantially higher resource requirements of the three-jet kernel, for which a floating-point implementation would lead to excessive pressure on FPGA resources and make timing closure significantly more difficult. A larger numerical discrepancy is therefore expected, due to quantization and scaling effects that accumulate over the more complex arithmetic structure of the kernel. In this case, the mean relative error is $4.104\times 10^{-1}\%$, with a standard deviation of $1.30\%$, while the maximum observed relative deviation reaches $11.40\%$.

To better characterize the tail of the error distribution, we also evaluated event-level percentiles using the maximum relative deviation among the four output components associated with each event. For this metric, the $P95$ and $P99$ percentiles are $0.719\%$ and $10.38\%$, respectively. These values indicate that the largest deviations are concentrated in a limited fraction of events, rather than being representative of the typical behavior of the kernel.

This broader spread in the relative error should be interpreted together with the absolute error metrics reported in Table~\ref{tab:numerical_validation_colour}. In particular, the mean absolute error remains at $4.682\times 10^{-6}$, although its standard deviation reaches $9.69\times 10^{-4}$, indicating a strongly non-uniform error distribution influenced by a limited number of larger deviations. The corresponding event-level absolute-error percentiles remain small, with $P95 = 1.46\times 10^{-8}$ and $P99 = 4.16\times 10^{-7}$.

For additional context, the main accumulator format used in the fixed-point three-jet kernel is \texttt{ap\_fixed<28,15>}, corresponding to an LSB of \(2^{-13} \approx 1.22\times 10^{-4}\) in the internal accumulation domain. This value provides a reference for the local numerical resolution of the dominant accumulation stage, but the reported absolute deviations should be interpreted as the cumulative result of quantization across the compact input representation, matrix--vector products, intermediate casts, and final reduction, rather than as the effect of a single quantization stage.

At the same time, the average absolute discrepancy remains small, even though the relative metric becomes much more sensitive in regions where the reference value is small. Taken together, these results indicate that the \texttt{ap\_fixed} FPGA implementation remains in reasonable numerical agreement with the double-precision reference for this benchmark, while enabling a substantially more resource-efficient realization of the three-jet kernel.

\begin{table*}[t]
\centering
\caption{Numerical validation of the colour-algebra kernels.}
\label{tab:numerical_validation_colour}
\resizebox{\textwidth}{!}{
\begin{tabular}{lcccccc}
\toprule
\textbf{Process} &
\makecell{\textbf{Mean rel.}\\\textbf{error [\%]}} &
\makecell{\textbf{Std. rel.}\\\textbf{error [\%]}} &
\makecell{\textbf{Max rel.}\\\textbf{error [\%]}} &
\makecell{\textbf{Mean abs.}\\\textbf{error}} &
\makecell{\textbf{Std. abs.}\\\textbf{error}} &
\makecell{\textbf{Max abs.}\\\textbf{error}} \\
\midrule
$t\bar{t} + 1$ jet & $5.432\times 10^{-5}$ & $9.16\times 10^{-5}$ & $6.678\times 10^{-4}$ & $5.438\times 10^{-9}$ & $2.50\times 10^{-8}$ & $4.071\times 10^{-7}$ \\
$t\bar{t} + 2$ jets & $8.573\times 10^{-4}$ & $1.65\times 10^{-3}$ & $6.602\times 10^{-2}$ & $2.372\times 10^{-10}$ & $9.64\times 10^{-10}$ & $6.278\times 10^{-8}$ \\
$t\bar{t} + 3$ jets & $4.104\times 10^{-1}$ & $1.30$ & $11.40$ & $4.682\times 10^{-6}$ & $9.69\times 10^{-4}$ & $2.472\times 10^{-1}$ \\
\bottomrule
\end{tabular}
}
\end{table*}

Overall, the numerical deviations introduced by the FPGA implementations remain small for the floating-point colour-algebra kernels. For the fixed-point three-jet colour-algebra kernel, the deviations are larger, as expected from the reduced-precision and scaled arithmetic used on the FPGA, but the absolute error metrics still indicate small average discrepancies for the considered benchmark. These results support the use of reduced-precision arithmetic as a practical compromise between numerical accuracy and hardware efficiency in the configurations studied here.

\subsubsection{Performance evaluation}
\label{sec:performance_colour}

For the \(gg \to t\bar{t}+X\) processes, the evaluation focuses on the
isolated colour-algebra kernels. These kernels represent only one stage of the
matrix-element calculation, although their computational cost increases
rapidly with jet multiplicity. Performance is characterised in terms of
per-event processing time because the kernels operate on precomputed
amplitudes. The reported values correspond to the mean over 10 independent
runs for samples of \(3\times 10^7\) events, and the associated variability is
quantified through the standard deviation. The measurements therefore isolate
the kernel-level cost of the colour reduction rather than the cost of the full
matrix-element evaluation or the complete event-generation workflow.
Host-side preparation, PCIe transfers, and other platform-specific
data-movement overheads are excluded from the measured interval. The reported
results consequently represent compute-only kernel execution times under the
stated measurement methodology.

In this context, an ``event'' corresponds to one invocation of the
colour-reduction kernel using precomputed \texttt{jamp} amplitudes as input,
rather than to the complete event-evaluation chain.

Table~\ref{tab:colour_performance} reports the average execution time per
event for the evaluated CPU, GPU, and FPGA implementations at different jet
multiplicities. The last column reports the ratio between the execution time
of each platform and that of the corresponding FPGA implementation,
\(\overline{T}_{\mathrm{platform}}/
\overline{T}_{\mathrm{FPGA}}\). Values greater than one therefore indicate
that the FPGA implementation is faster, whereas values below one indicate
that the compared platform is faster.

\begin{table*}[t]
\centering
\caption{Average execution time per event for the colour-algebra kernels.
The reported values correspond to the isolated colour-reduction kernel
operating on precomputed amplitudes. The last column reports
\(\overline{T}_{\mathrm{platform}}/
\overline{T}_{\mathrm{FPGA}}\) for each jet multiplicity.}
\label{tab:colour_performance}
\resizebox{\textwidth}{!}{
\begin{tabular}{lcccc}
\toprule
\textbf{Process} &
\textbf{Platform} &
\makecell{\textbf{Average time/event}\\\textbf{[ns]}} &
\makecell{\textbf{Std.}\\\textbf{[ns]}} &
\makecell{\textbf{Relative to}\\\textbf{FPGA}} \\
\midrule
$t\bar{t}+1$ jet
  & CPU AMD EPYC 9474F      & 33.46    & 0.06 & 2.57 \\
  & CPU Intel i7-13700      & 40.52    & 0.03 & 3.11 \\
  & GPU NVIDIA RTX 6000     & 54.75    & 0.02 & 4.21 \\
  & GPU NVIDIA H100         & 4.74     & 0.01 & 0.36 \\
  & FPGA Alveo U250 (1 CU)  & 13.00    & 0.03 & 1.00 \\
\midrule
$t\bar{t}+2$ jets
  & CPU AMD EPYC 9474F      & 379.68   & 7.84 & 23.73 \\
  & CPU Intel i7-13700      & 560.09   & 6.47 & 35.01 \\
  & GPU NVIDIA RTX 6000     & 707.73   & 4.89 & 44.23 \\
  & GPU NVIDIA H100         & 155.99   & 0.07 & 9.75 \\
  & FPGA Alveo U250 (1 CU)  & 16.00    & 0.04 & 1.00 \\
\midrule
$t\bar{t}+3$ jets
  & CPU AMD EPYC 9474F      & 20622.44 & 51.98 & 389.10 \\
  & CPU Intel i7-13700      & 29697.54 & 71.48 & 560.32 \\
  & GPU NVIDIA RTX 6000     & 12988.43 & 76.77 & 245.06 \\
  & GPU NVIDIA H100         & 4522.31  & 2.73  & 85.33 \\
  & FPGA Alveo U250 (1 CU)  & 53.00    & 0.07 & 1.00 \\
\bottomrule
\end{tabular}
}
\end{table*}

For the \(t\bar{t}+1\)-jet kernel, the FPGA achieves speedups of approximately
\(2.57\times\) and \(3.11\times\) over the AMD EPYC 9474F and Intel
i7-13700 CPU baselines, respectively, and \(4.21\times\) over the
RTX 6000. The H100 is faster in this simplest case, requiring
\(4.74\,\mathrm{ns}\) per event compared with \(13.00\,\mathrm{ns}\) for
the FPGA.

For \(t\bar{t}+2\) jets, the FPGA provides speedups of approximately
\(23.7\times\) and \(35.0\times\) over the two CPU baselines, respectively,
and \(44.2\times\) and \(9.75\times\) over the RTX 6000 and H100. For the
most demanding \(t\bar{t}+3\)-jet kernel, the corresponding speedups relative
to the AMD EPYC 9474F, Intel i7-13700, RTX 6000, and H100 are approximately
\(389\times\), \(560\times\), \(245\times\), and \(85.3\times\),
respectively.

Across the evaluated implementations, the measured advantage of the FPGA
increases with jet multiplicity. The GPU results also depend strongly on the
complexity of the target kernel. The H100 achieves lower per-event execution
times than both CPU baselines in all three cases, while the RTX 6000 becomes
faster than the CPU baselines only for the three-jet kernel. The FPGA achieves
the lowest measured per-event execution time for the two- and three-jet
kernels, whose regular arithmetic structure and dataflow organisation are
well suited to a deeply pipelined implementation.

At the same time, these kernel-level speedups do not translate directly into
equivalent speedups of the complete workflow. In the profiled
\(t\bar{t}+2\)-jet case, the colour-reduction stage represents slightly more
than \(2\%\) of the total execution time. For the higher-multiplicity
\(t\bar{t}+3\)-jet process, CPU profiling shows that colour summation accounts
for approximately \(7\%\) of the total execution time. The increase from
slightly more than \(2\%\) to approximately \(7\%\) provides additional
quantitative evidence of the growing computational relevance of the
colour-algebra stage with increasing multiplicity.

Consequently, even a substantial acceleration of this isolated stage should
not be interpreted as an equivalent end-to-end workflow speedup. The values
reported in Table~\ref{tab:colour_performance} should therefore be interpreted
as kernel-level speedups rather than as speedups of the full matrix-element
evaluation or the complete event-generation chain.

Overall, these results provide a promising starting point and demonstrate the
potential of FPGA acceleration for highly regular colour-algebra kernels.
Achieving a significant workflow-level benefit would require extending the
accelerated region to a larger fraction of the computation and evaluating the
effects of host--FPGA data movement, kernel orchestration, and integration
with the remaining stages of the event-generation workflow.

\subsubsection{Throughput efficiency analysis}
\label{sec:throughput_efficiency_colour}

Table~\ref{tab:fpga_impl_summary} summarizes the main post-implementation timing characteristics of the evaluated FPGA kernels, including target and achieved clock frequencies, achieved initiation interval, pipeline latency, and worst negative slack.

These results confirm that the one-jet and two-jet colour kernels achieve \(\mathrm{II}=1\), while the more demanding three-jet kernel operates at \(\mathrm{II}=4\), consistent with the practical throughput bound discussed below.

From an architectural point of view, the absolute ideal limit for the colour-algebra kernels would also correspond to 10~ns/event at 100~MHz, i.e.\ one event processed per clock cycle with \(\mathrm{II}=1\). For the more complex kernels, however, this limit is not achievable in practice within the available FPGA resources while maintaining timing closure.

For this reason, the implemented kernels adopt a more resource-aware organization. In the three-jet case, the computation is decomposed into a multi-stage dataflow architecture in which the effective lower bound is determined by the input and unpacking stages, leading to a realizable ideal throughput of one event every four clock cycles. At 100~MHz, this corresponds to 40~ns/event. Thus, for the colour-algebra kernels it is useful to distinguish between the absolute ideal limit of 10~ns/event and the practical ideal bound imposed by the implemented architecture.

Table~\ref{tab:colour_ideal} compares the measured per-event execution time with both the absolute ideal bound and the practical ideal bound of the implemented architecture. For the one-jet and two-jet cases, the measured times remain reasonably close to the absolute 10~ns limit, reaching about 76.9\% and 62.5\% efficiency, respectively. For the three-jet implementation, the measured execution time is 53~ns/event, corresponding to 18.9\% of the absolute ideal throughput, but 75.5\% of the practical ideal bound of the adopted 4-cycle design.

\begin{table}[t]
\centering
\caption{Measured performance of the colour-algebra kernels compared with absolute and practical ideal limits.}
\label{tab:colour_ideal}
\resizebox{\columnwidth}{!}{%
\begin{tabular}{lccccc}
\toprule
\textbf{Process} &
\makecell{\textbf{Measured}\\\textbf{[ns]}} &
\makecell{\textbf{Abs. Ideal}\\\textbf{[ns]}} & 
\makecell{\textbf{Practical Ideal}\\\textbf{[ns]}} & 
\makecell{\textbf{Eff. (abs.)}\\\textbf{[\%]}} & 
\makecell{\textbf{Eff. (practical)}\\\textbf{[\%]}}  \\
\midrule
$t\bar{t} + 1$ jet & 13 & 10 & 10 & 76.9 & 76.9 \\
$t\bar{t} + 2$ jets & 16 & 10 & 10 & 62.5 & 62.5 \\
$t\bar{t} + 3$ jets & 53 & 10 & 40 & 18.9 & 75.5 \\
\bottomrule
\end{tabular}%
}
\end{table}

The gap between measured and ideal performance reflects the non-negligible cost of data unpacking, memory accesses, internal reductions, and inter-stage synchronization. For the three-jet case in particular, the implemented design deliberately relaxes the absolute one-event-per-cycle target in order to satisfy resource and timing constraints, while still remaining relatively close to the best achievable bound of the adopted architecture.

Overall, this analysis shows that the measured performance of the colour-algebra kernels remains reasonably close to the limits implied by the adopted hardware organization. At the same time, it highlights that the gap with respect to the ideal limit grows with the complexity of the target kernel.

\subsubsection{Resource utilization and scalability}
\label{sec:resources_scalability_colour}

Table~\ref{tab:resource_colour} summarizes the FPGA resource utilization for the colour-algebra kernels. The one- and two-jet implementations use floating-point arithmetic, while the three-jet implementation adopts a fixed-point representation in order to reduce resource pressure and enable timing closure for the substantially larger colour basis.

\begin{table*}[t]
\centering
\caption{FPGA resource utilization for the colour-algebra kernels.}
\label{tab:resource_colour}
\resizebox{\textwidth}{!}{
\begin{tabular}{lccccc}
\toprule
\textbf{Process} & \textbf{Numerical format} & \textbf{LUTs} & \textbf{FFs} & \textbf{DSPs} & \textbf{BRAM} \\
\midrule
$t\bar{t} + 1$ jet & Float & 66159 (3.75\%) & 86472 (2.36\%) & 912 (7.34\%) & 64 (1.17\%) \\
$t\bar{t} + 2$ jets & Float & 792986 (45.10\%) & 617610 (16.87\%) & 9288 (74.75\%) & 90 (1.65\%) \\
$t\bar{t} + 3$ jets & Fixed-point & 214246 (12.17\%) & 57076 (1.56\%) & 4734 (38.10\%) & 914 (16.79\%) \\
\bottomrule
\end{tabular}
}
\end{table*}

The resource-utilization trends for the colour-algebra kernels reflect both the rapid growth in computational complexity with jet multiplicity and the strong impact of the adopted numerical representation. In particular, the larger colour bases in the two-jet and three-jet cases lead to a substantial increase in hardware cost with respect to the one-jet kernel. However, the adopted numerical representation also plays a major role in determining the final resource footprint. For this reason, the two-jet and three-jet implementations should not be interpreted as a like-for-like comparison of hardware cost at fixed numerical precision, but rather as an illustration of how numerical representation reshapes the resource footprint of increasingly complex kernels.

For the floating-point implementations, DSP usage dominates, reflecting the cost of complex arithmetic in the matrix-contraction stage. This effect is especially pronounced in the \(t\bar{t}+2\) jets case, where the combination of a larger colour basis and floating-point arithmetic drives the utilization to 45.1\% of LUTs and 74.8\% of DSPs, placing the design close to the practical resource limits of the device.

This strong resource pressure motivates the transition to fixed-point arithmetic in the three-jet implementation. Despite the larger problem size, the use of a more compact numerical representation makes it possible to reduce LUT and DSP usage with respect to a floating-point design of comparable complexity, while enabling timing closure. This comes at the expense of a substantial increase in BRAM usage, which reflects the larger storage requirements of the adopted architecture.

These results show that the achievable scalability of the colour-algebra kernels is determined jointly by the complexity of the colour basis and by the numerical representation used to implement it. In particular, the use of fixed-point arithmetic in the three-jet kernel makes it possible to realize an FPGA implementation for a substantially larger problem than would be practical with a purely floating-point design.

Overall, both the complexity of the target physics process and the chosen numerical representation play a central role in determining the scalability of FPGA-based implementations. In practice, resource usage and timing closure jointly constrain the achievable level of parallelism, particularly for the more demanding colour-algebra kernels.

\section{Conclusion}
In this work, we have presented an FPGA-based approach to the acceleration of
matrix-element calculations in high-energy physics, considering two
complementary scenarios: an implementation of the phase-space-generation and
matrix-element-evaluation stages for the benchmark process
\(e^+e^- \to \mu^+\mu^-\), and a kernel-level acceleration study of the
colour algebra for \(gg \to t\bar{t}+X\) processes with increasing jet
multiplicity. The proposed designs rely on streaming architectures and deeply
pipelined execution to exploit fine-grained parallelism and sustain
high-throughput processing.

For the \(e^+e^- \to \mu^+\mu^-\) benchmark, the evaluated computational
stages were mapped onto the FPGA. Under the compute-only timing methodology
and execution configurations considered in this work, the 8-CU FPGA
implementation achieves speedups of approximately \(95.5\times\) over the
Intel i7-13700 CPU baseline, \(86.6\times\) over the RTX 3050,
\(10.1\times\) over the RTX 6000, and \(6.35\times\) over the H100. The
numerical comparison against the double-precision CPU reference indicates
that the reduced-precision FPGA implementation remains in close numerical
agreement with the reference for this benchmark. These results demonstrate
the potential of a deeply pipelined FPGA implementation for this particular
process and measurement scope, rather than establishing general
workflow-level competitiveness.

For the more complex \(gg \to t\bar{t}+X\) processes, the study focused on
the colour-algebra stage as a representative structured kernel within the
\mgfive\ workflow. The reported speedups correspond to the isolated
colour-reduction kernel operating on precomputed amplitudes and should not be
interpreted as speedups of the full matrix-element evaluation or the complete
event-generation workflow. Within this scope, the measured FPGA advantage
increases strongly with jet multiplicity. For the \(t\bar{t}+3\)-jet kernel,
the FPGA implementation achieves speedups of approximately \(389\times\),
\(560\times\), \(245\times\), and \(85\times\) relative to the AMD EPYC
9474F, Intel i7-13700, RTX 6000, and H100 baselines, respectively.

In the profiled \(t\bar{t}+2\)-jet workflow, the colour-reduction stage
represents slightly more than \(2\%\) of the total runtime. For the
higher-multiplicity \(t\bar{t}+3\)-jet process, CPU profiling shows that
colour summation accounts for approximately \(7\%\) of the total execution
time. The increase from slightly more than \(2\%\) to approximately \(7\%\),
together with the rapid increase in colour-basis size, provides further
evidence of the growing computational relevance of the colour-algebra stage
with increasing multiplicity.

Nevertheless, even a substantial acceleration of the isolated colour kernel
should not be interpreted as an equivalent end-to-end workflow speedup. The
colour-algebra results should therefore be regarded as a promising starting
point and as evidence of the potential of FPGA acceleration for highly regular
computational kernels, rather than as a demonstration of a practical
acceleration strategy for the complete \mgfive\ workflow.

A central outcome of this work is the strong impact of numerical
representation on performance, scalability, and resource utilisation.
Floating-point arithmetic provides a straightforward implementation for
moderate problem sizes but introduces substantial resource pressure as the
complexity of the colour basis grows. Fixed-point arithmetic was therefore
adopted for the most demanding kernels to reduce this pressure and support
larger implementations, at the cost of increased numerical deviation. For
the benchmarks considered here, the retained formats provide acceptable
agreement with the corresponding software references under the numerical
validation methodology described in this work.

The present study also has important limitations. FPGA scalability is
constrained by resource availability, routing complexity, timing-closure
pressure, and memory-system effects, which lead to sublinear scaling at the
largest parallel configurations. In addition, the reported measurements
exclude host-side preparation, PCIe transfers, buffer-management costs, and
CPU--FPGA orchestration. These overheads can be significant in a heterogeneous
workflow and may reduce or eliminate the performance advantage observed for
an isolated accelerator kernel. The current results therefore do not
establish that FPGAs are generally competitive for matrix-element calculations
or complete Monte Carlo event-generation workflows.

Future work will focus on evaluating more integrated implementations in which
a larger fraction of the event-generation workflow is executed on the
accelerator. This includes measuring CPU--FPGA handoff and data-movement
costs, investigating batching and communication--computation overlap, and
exploring mixed-precision arithmetic, improved memory organisations, and
multi-FPGA execution. Such studies are required to determine whether the
kernel-level advantages observed here can translate into significant
end-to-end benefits for realistic production workflows.

Overall, this work demonstrates substantial compute-only acceleration for a
specific full matrix-element benchmark and for selected structured
colour-algebra kernels under the evaluated conditions. These results establish
the technical potential of FPGA acceleration for suitable components of
matrix-element calculations, while further end-to-end evaluation is required
before conclusions can be drawn regarding its general viability or
competitiveness in heterogeneous Monte Carlo event-generation systems.

\section*{Statements and Declarations}

\subsection*{Declaration of interests}
The authors declare that they have no known competing financial interests or personal relationships that could have appeared to influence the work reported in this paper.

\subsection*{Funding}
This research was funded by the Ministerio de Ciencia, Innovaci\'on y Universidades, with support from NextGenerationEU and the Plan de Recuperaci\'on, Transformaci\'on y Resiliencia, under project TED2021-130852B-I00.

This work and the authors are also partially supported by the European Research Council (ERC) under the European Union's Horizon Europe research and innovation programme through the grant INTREPID (grant agreement No.\ 101115353). Views and opinions expressed are, however, those of the author(s) only and do not necessarily reflect those of the European Union or the European Research Council Executive Agency. Neither the European Union nor the granting authority can be held responsible for them.

\subsection*{Author contributions}
H.~Guti\'{e}rrez Arance conceived and carried out the work, including the
hardware and software implementation, validation, analysis, visualisation, and
manuscript drafting.
C.~Vico Villalba contributed expertise in Monte Carlo event generation and
reviewed the manuscript.
S.~Folgueras contributed to project administration, supported the work through
funding acquisition, and reviewed the manuscript.
F.~Carri\'{o} contributed expertise on the AMD hardware platform and
reviewed the manuscript.
A.~Valero and L.~Fiorini contributed through technical discussion, project
administration, and manuscript review.
P.~Leguina L\'{o}pez and F.~Herv\'{a}s \'{A}lvarez contributed through investigation, technical discussion, and manuscript review.
A.~Oyanguren contributed to project administration and funding support.

\subsection*{Data availability}
The \LaTeX\ sources, figure files, FPGA kernels, and supporting analysis scripts are available at the project repositories.\footnote{\url{https://gitlab.cern.ch/hgutierr/epem_mupmum_fixed_hls/-/releases/paper-v1.0}}
\footnote{\url{https://gitlab.cern.ch/hgutierr/color_gg_ttg_hls/-/releases/paper-v1.0}}
\footnote{\url{https://gitlab.cern.ch/hgutierr/color_gg_ttgg_hls/-/releases/paper-v1.0}}
\footnote{\url{https://gitlab.cern.ch/hgutierr/color_gg_ttggg_hls/-/releases/paper-v1.0}}
Additional data are available from the corresponding author upon reasonable request.

\subsection*{Declaration of generative AI and AI-assisted technologies in the manuscript preparation process}
During the preparation of this work, the authors used ChatGPT (OpenAI) to assist with language refinement, text editing, and improvements in clarity in parts of the manuscript. After using this tool, the authors reviewed and edited the content as needed and take full responsibility for the content of the published article.
%\bibliographystyle{IEEEtran}
%\bibliography{references}
\printbibliography
\end{document}